\documentclass[fleqn,usenatbib]{mnras}

\usepackage{newtxtext,newtxmath}

\usepackage[T1]{fontenc}

\DeclareRobustCommand{\VAN}[3]{#2}
\let\VANthebibliography\thebibliography
\def\thebibliography{\DeclareRobustCommand{\VAN}[3]{##3}\VANthebibliography}

\usepackage{graphicx}	
\usepackage{amsmath}	
\usepackage{mathtools}
\usepackage{pdflscape}
\usepackage{natbib}
\usepackage{bm}
\usepackage{longtable}
\usepackage{hyperref}
\usepackage{rotating}
\usepackage{subcaption}
\usepackage{tablefootnote}
\usepackage{booktabs}

\newcommand{\bd}{BD+46$^{\circ}$442}
\newcommand{\angstrom}{\AA}
\newcommand{\teff}{T_\mathrm{eff}}

\newcommand{\myr}{M$_\odot\,\text{yr}^{-1}$}
\newcommand{\kms}{km\,s$^{-1}$}
\newcommand{\halpha}{H$\alpha$}
\newcommand{\hbeta}{H$\beta$}
\newcommand{\hgamma}{H$\gamma$}
\newcommand{\hdelta}{H$\delta$}

\newcommand{\iras}{IRAS19135+3937}

\title[The diversity of post-AGB binaries with jets.]{Jet parameters for a diverse sample of jet-launching post-AGB binaries}

\author[D. Bollen et al.]{Dylan Bollen,$^{1,2,3}$\thanks{E-mail: dylan.bollen@kuleuven.be (DB)}
Devika Kamath,$^{2,3}$
Hans Van Winckel,$^{1}$
Orsola De Marco$^{2,3}$
\newauthor
and Mark Wardle$^{2,3}$
\\
$^{1}$Institute of Astronomy, KU Leuven, Celestijnenlaan 200D, B-3001 Leuven, Belgium\\
$^{2}$Department of Physics \& Astronomy, Macquarie University, Sydney, NSW 2109, Australia\\
$^{3}$Astronomy, Astrophysics and Astrophotonics Research Centre, Macquarie University, Sydney, NSW 2109, Australia
}

\date{Accepted XXX. Received YYY; in original form ZZZ}

\pubyear{2020}

\begin{document}
\label{firstpage}
\pagerange{\pageref{firstpage}--\pageref{lastpage}}
\maketitle

\begin{abstract}
Jets are a commonly observed phenomenon in post-asymptotic giant branch (post-AGB) binaries. Due to the orbital motion of the binary, the jet causes variable absorption in the Balmer profiles. In previous work, we have developed spatio-kinematic and radiative transfer models to reproduce the observed Balmer line variability and derive the spatio-kinematic structure of the jet and its mass-loss rate. Here, we apply our jet model to five post-AGB binaries with distinct \halpha\,line variability and diverse orbital properties. Our models fit the \halpha\, line variations very well. We estimate jet mass-loss rates between $10^{-8}\,$\myr and $10^{-4}\,$\myr, from which we deduce accretion rates onto the companion between $10^{-7}\,$\myr and $10^{-3}\,$\myr. These accretion rates are somewhat higher than can be comfortably explained with reasonable sources of accretion, but we argue that the circumbinary disc in these systems is most-likely the source feeding the accretion, although accretion from the post-AGB star cannot be ruled out. The diversity of the variability in the five objects is due to their wide ejection cones combined with a range of viewing angles, rather than inherent differences between the objects. The nature of the observations does not let us easily distinguish which jet launching model (stellar jet, disc wind, or X-wind) should be favoured. In conclusion, we show that our jet model includes the physical parameters to successfully reproduce the \halpha\,line variations and retrieve the structure and mass-loss rates of the jet for all five objects that are representative of the diverse sample of Galactic post-AGB binaries. 
\end{abstract}




\section{Introduction}\label{sec:intro}

The post-AGB phase of low to intermediate mass stars is a short-lived phase during which the star changes dramatically \citep{vanwinckel03}. This evolutionary phase becomes more complex if the post-AGB star is part of a binary system. Binary interactions during the post-AGB and preceding phases can fundamentally change the stellar properties and evolution of both stars.

Many of these post-AGB binary systems are orbited by large, circumbinary Keplerian discs of gas and dust \citep{bujarrabal13a, bujarrabal18, vanaarle11, deruyter06, vanwinckel09, kamath14a, kamath15a}. These post-AGB binaries with Keplerian discs have periods in the range $\sim100-1200$~days \citep{vanwinckel09, oomen18}. The discs around these binary systems are likely created by a combination of high mass-loss rate during the AGB phase and the focusing action of a binary companion. This focusing action can concentrate the mass loss on the orbital plane and donate angular momentum, thus creating a circumbinary, Keplerian disc \citep{shu79, pejcha16, macleod18, hubova19, bermudez20}. Their characteristics hold clues to a number of open questions in stellar astrophysics, such as the nature of consequence of binary interactions between giants and compact stars, something that gives rise to scores of evolved binaries such as cataclysmic variables, or the progenitors of type Ia supernovae \citep{DeMarco2017}.

Strong binary interactions on the AGB almost invariably lead to a common envelope in-spiral and merger, or to a close binary \citep{DeMarco2015}. As such, the existence of post-AGB binaries, particularly those with periods $\sim$100 days, is puzzling: these periods are much larger than post-common envelope periods, but too short to have avoided a common envelope interaction altogether. Neither do we understand the non-zero eccentricities of these binaries, since the tidal forces working on the binary system are expected to circularise the orbit \citep{zahn77}. Whether the period distribution of the post-AGB binary population is due to their having mass ratios near unity at the time of common envelope \citep{oomen18,iaconi19}, or whether the period is widened {\it after} the binary interaction on the AGB, is not clear. Neither is it clear whether these wide binaries will ever form a visible planetary nebula, and hence whether they will become the elusive, wider binaries central star population \citep{Douchin2015,JOnes2017}.

A peculiar observational signature of post-AGB stars in binary systems is an under-abundance of refractory elements in the spectra of their stellar atmospheres \citep[e.g.][]{vanwinckel95,vanwinckel97, maas02}. This depletion pattern is only observed in post-AGB stars with a circumbinary disc \citep{gezer15, kamath19}. It is currently understood that the pattern is caused by the re-accretion of gas from the circumbinary disc, since the re-accreted gas becomes poor in refractory elements after dust formation in the disc \citep[][and references therein]{oomen19}. 

\cite{kamath16} discovered that the fainter post-AGB binaries with circumbinary discs are actually post-red giant branch (post-RGB) stars that have evolved off the RGB instead of the AGB. A subset of the entire post-AGB/post-RGB binary sample populates the high-luminosity end of the population II cepheid instability strip \citep{wallerstein02}. These pulsating giants are referred to as RV Tauri stars with dust. The photometric variability of these stars have typical pulsation periods between 20 and 150 days and a characteristic light curve with an alternation between deep and shallow minima \citep{wallerstein02, manick17, manick19}. Some RV Tauri variables show a secondary long-term variability with a period between 600 and 2600 days \citep{pollard96}, known as photometric type RVb variables. The variability has been linked to the binary nature of the system and the presence of a circumbinary disc \citep{waelkens95, vanwinckel99,manick17}. When the inclination angle of the binary system is high enough (about $60\degr$ or higher), the post-AGB star will be obscured partially by the circumbinary disc during inferior conjunction, when the post-AGB star is closest to the observer. This causes a periodic photometric variability which is synchronous with the orbital period of the binary system due to variable line-of-sight extinction.

\cite{gorlova12, gorlova15} performed time-series analyses of high-resolution optical spectra of post-AGB binaries \bd\, and \iras, and found interesting variations in the \halpha\,line. They found that the \halpha\,line has a broad, blue-shifted absorption feature during superior conjunction, when the post-AGB star is located behind the companion. The origin of this orbital-phase dependent line variation is the presence of a jet launched from an accretion disc around the companion star \citep{gorlova12,bollen17}. During superior conjunction, the jet blocks the photospheric light from the post-AGB star. Hence, the hydrogen atoms in the jet scatter the photospheric light from the post-AGB star out of the line-of-sight, causing this particular absorption feature. Due to the orbital motion of the binary, the light of the post-AGB star shines through different parts of the jet over time, thus allowing for a tomography of the jet.

In \cite{bollen19}, we built a spatio-kinematic model of the jet, from which
synthetic spectra were created and fitted to the observations. This provided important jet parameters, such as the jet angle, velocity, and scaled density structure. We found that the jet in \bd\, has a low density inner cavity. This jet cavity can be caused by the launching geometry of the jet. In the disc-wind model by \cite{blandford82}, the disc matter is launched from the disc surface. The launching region extends outwards from the inner disc rim and the matter is launched at an angle larger than $30\degr$.This results in an inner jet region devoid of matter, similar to what we observed. 

In \cite{bollen20}, we improved our spatio-kinematic model and built an additional radiative transfer model, from which the absolute jet density structure was estimated. The results of both models allowed us to calculate the jet mass-loss rates and mass-accretion rates onto the companion. We found high mass-loss rates in the order of $10^{-7}-10^{-5}\,$\myr, which can be explained by re-accretion from the circumbinary disc onto the binary system. This is in line with the observed depletion pattern in post-AGB stars with circumbinary discs. In \cite{bollen20}, we tested our spatio-kinematic and radiative transfer models using \bd\, and \iras. The dynamic spectra and orbital parameters of these two objects were very similar (e.g., orbital periods of $140.8$ days and $127\,$ days, eccentricities of $0.08$ and $0.13$, and orbital separations of $0.63\,$AU and $0.55\,$AU, for \bd\, and \iras, respectively). 

In this work, we apply our spatio-kinematic and radiative transfer model to five objects with diverse dynamic \halpha\,spectra and orbital parameters, different from the two objects studied in our previous work. We aim to show that our models contain sufficient physical parameters to successfully reproduce the observed \halpha\,line variations for a diverse sample of jet-creating post-AGB binaries.

This paper is structured as follows: we introduce the spectroscopic data and the post-AGB binaries in our sample in Section~\ref{sec:obs}. We explain our methods for the modelling of these systems in Section~\ref{sec:methods}. The results of the spatio-kinematic and radiative transfer modelling are provided in Section~\ref{sec:results}. We discuss these results in Section~\ref{sec:discussion}, where we focus on the nature of the jet and the mass-transfer in the system. We end this paper with a summary and conclusions in Section~\ref{sec:conclusions}. 


\section{Observations and sample selection}\label{sec:obs}


\subsection{Spectroscopic data}\label{ssec:spec}

\begin{table} 
    \begin{center}
    \caption{Listed are the number of observations $N_\text{obs}$ and the number of observations selected for this study $N_\text{used}$, the total length of the observations $\Delta T_\text{obs}$, the orbital period of the binary system $P$, the amount of orbital cycles covered by the observations $\Delta P_\text{obs}$, and the average signal-to noise in \halpha\,for each object (S/N)$_\text{avg}$.}
    \label{tab:specdata}
    \begin{tabular}{l c c c c c c}
        \hline
        Object     & $N_\text{obs}$ & $N_\text{used}$ & $\Delta T_\text{obs}$ & $P$ & $\Delta P_\text{obs}$ &(S/N)$_\text{avg}$ \\
                   &                &                & days                  & days            & cycles                &  \\
        \hline
        89\,Her    & 210            &      26        & 1140                  & 289.1           & 3.9                   & $\sim125$\\
        EP\,Lyr    & 91             &      30        & 2923                  & 1151            & 2.5                   & $\sim45$\\
        HD\,46703  & 86             &      34        & 1624                  & 597.4           & 2.7                   & $\sim60$\\
        HP\,Lyr    & 91             &      40        & 1670                  & 1818            & 0.92                  & $\sim55$\\
        TW\,Cam    & 197            &      42        & 1634                  & 662             & 2.5                   & $\sim50$\\
        \hline
    \end{tabular}
    \end{center}
\end{table}

\begin{figure*}
  \centering
    \begin{subfigure}[b]{.4\linewidth}
        \centering\large 
          \includegraphics[width=.9\linewidth]{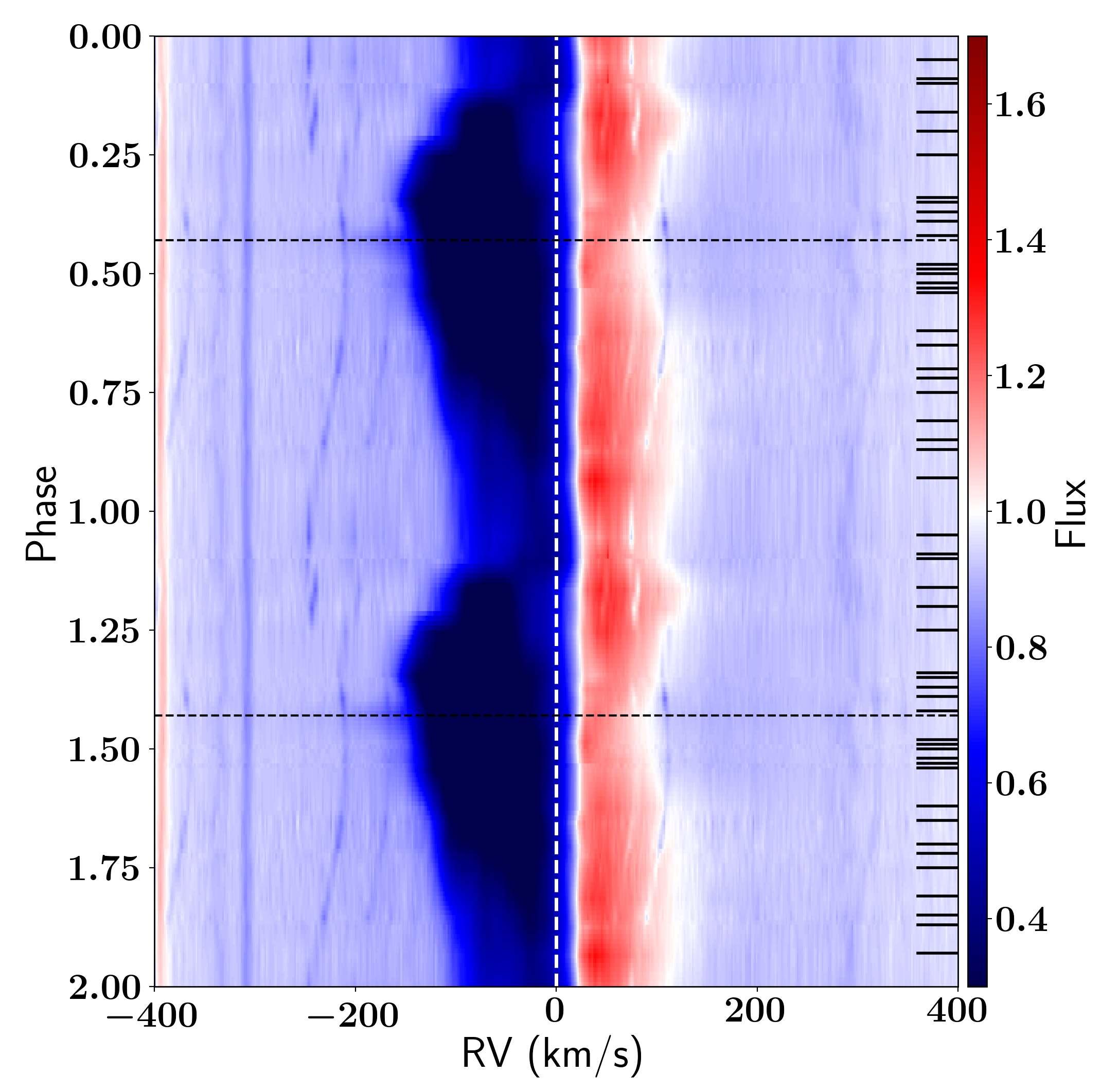}
        \caption{89\,Her}\label{fig:89_dynspec}
    \end{subfigure}%
    \begin{subfigure}[b]{.4\linewidth}
        \centering\large 
          \includegraphics[width=.9\linewidth]{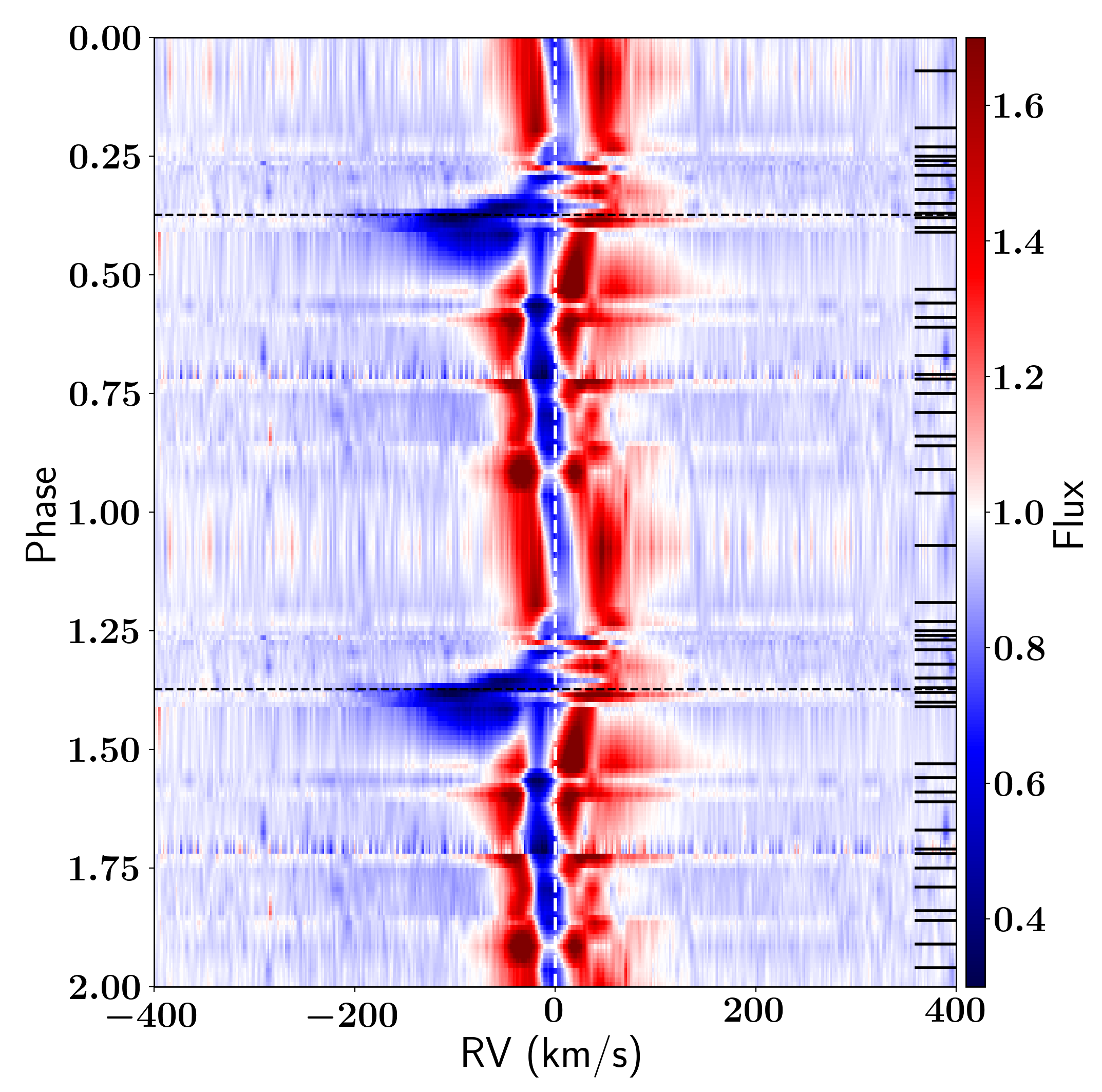}
        \caption{EP Lyr}\label{fig:ep_dynspec}
    \end{subfigure}%
    
    \begin{subfigure}[b]{.4\linewidth}
        \centering\large 
          \includegraphics[width=.9\linewidth]{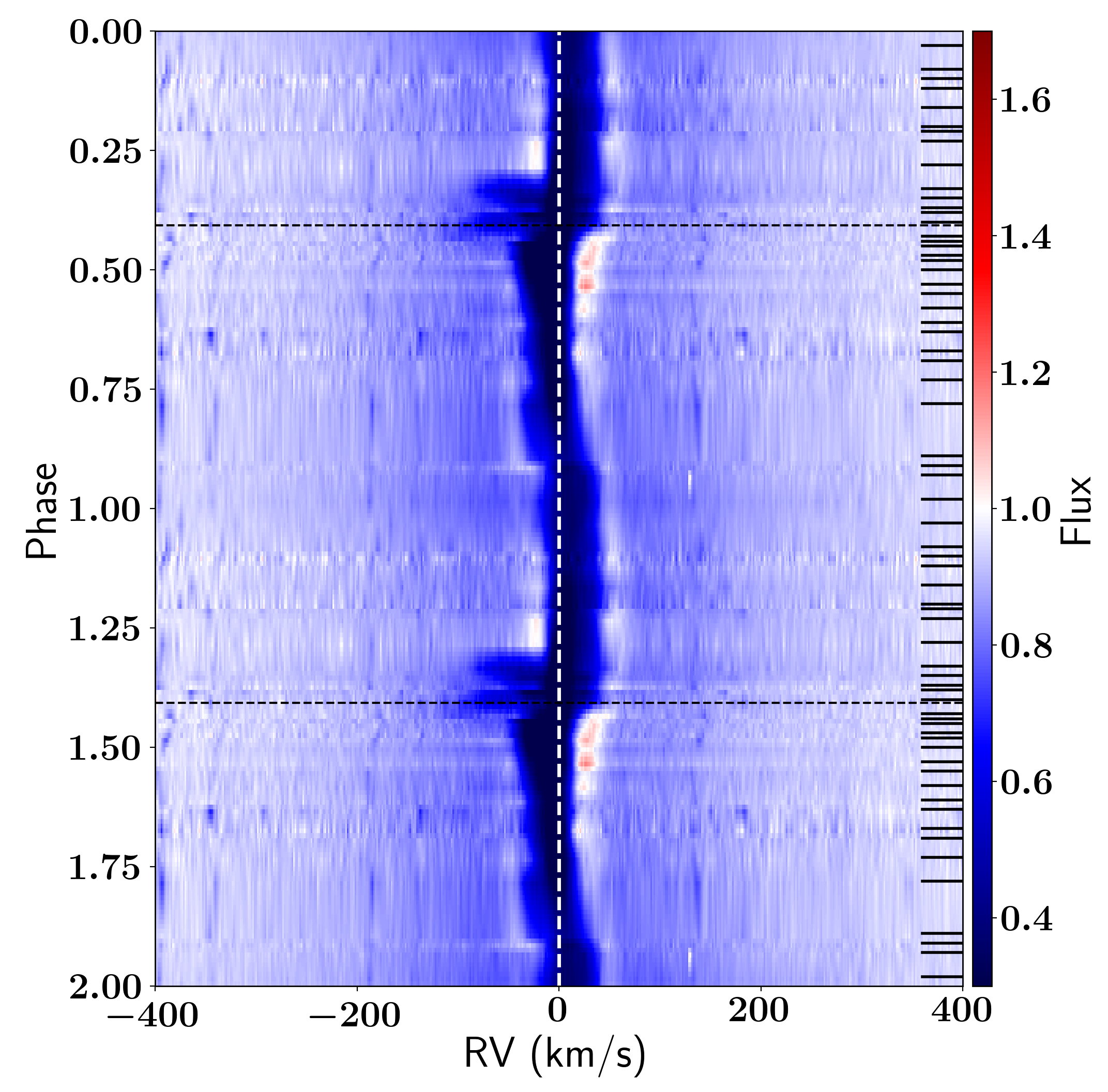}
        \caption{HD 46703}\label{fig:hd_dynspec}
    \end{subfigure}%
    \begin{subfigure}[b]{.4\linewidth}
        \centering\large 
          \includegraphics[width=.9\linewidth]{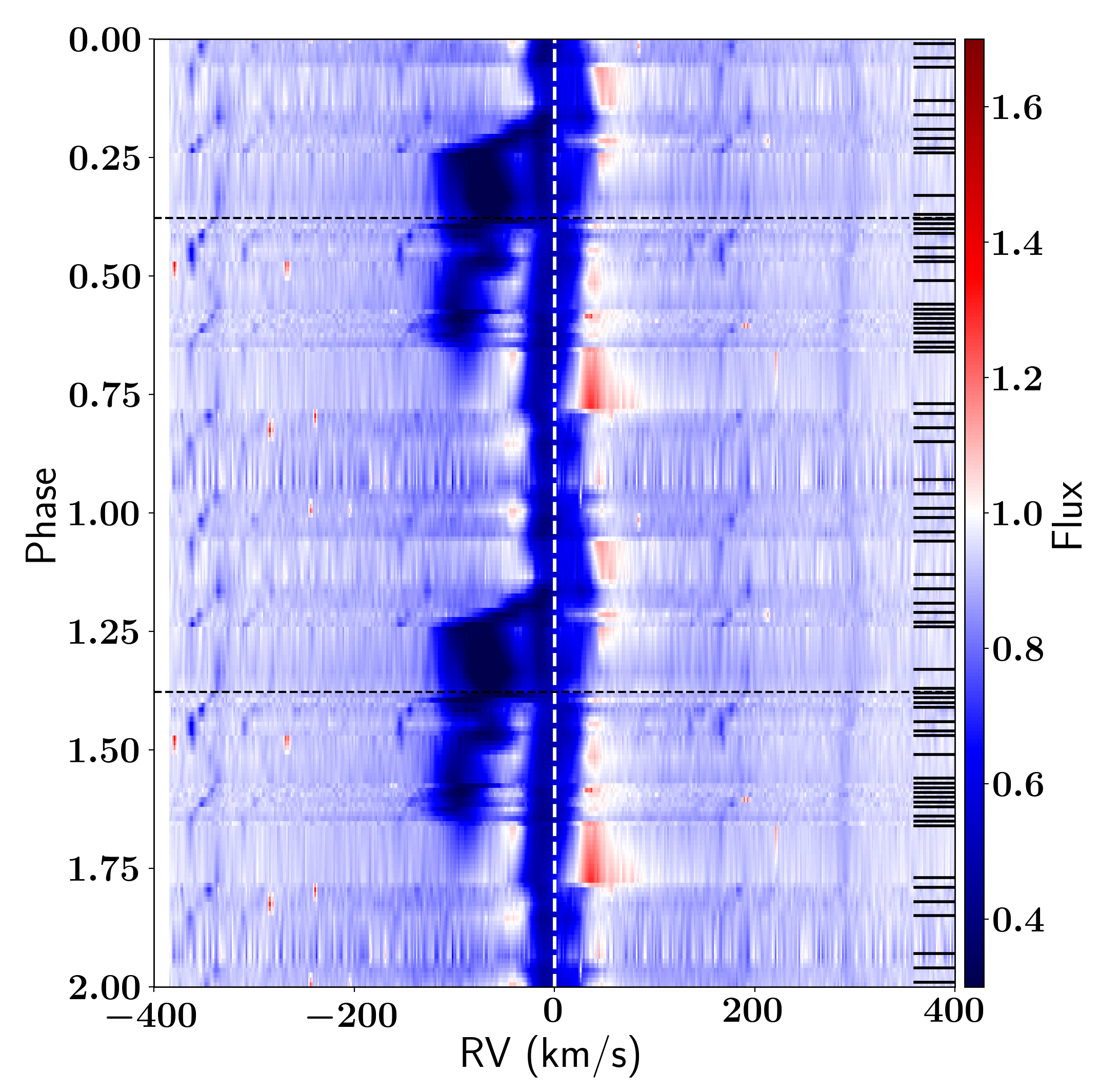}
        \caption{HP Lyr}\label{fig:hp_dynspec}
    \end{subfigure}%
    
    \begin{subfigure}[b]{.4\linewidth}
        \centering\large 
          \includegraphics[width=.9\linewidth]{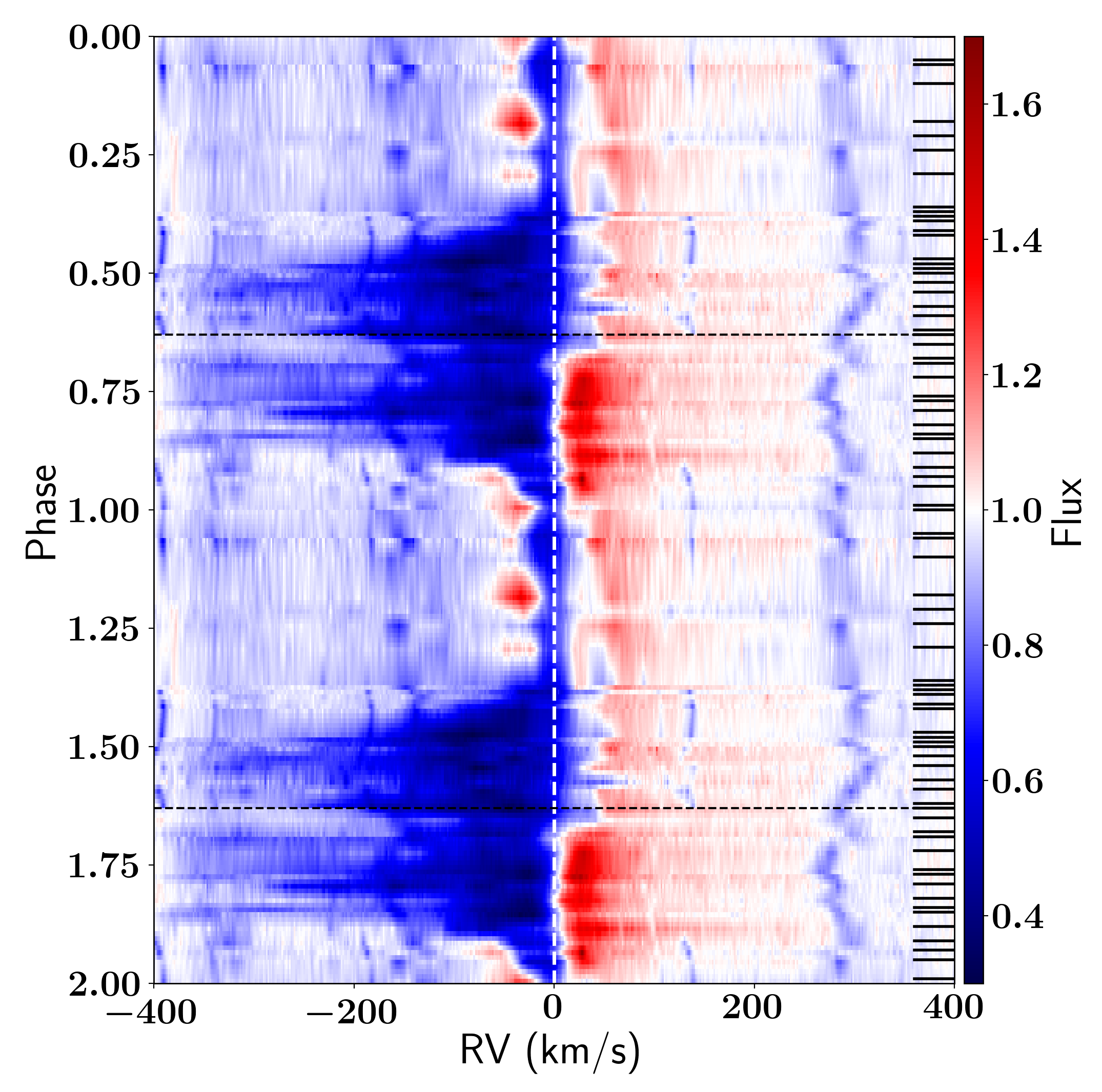}
        \caption{TW Cam}\label{fig:tw_dynspec}
    \end{subfigure}%
    \caption{The dynamic spectra for the \halpha\,line of the five objects in our sample. The spectra are shown as a function of orbital phase. The zero point of the radial velocity is set to the systemic velocity. The black dashed line indicates the phase of superior conjunction. The black horizontal dashes indicate the orbital phases during which a spectrum is taken. To guide the eye, we display the folded spectra twice.} \label{fig:dynspec}
\end{figure*}

The five objects in this sample (i.e., 89 Her, EP Lyr, HD 46703, HP Lyr, and TW Cam) are part of a long-term radial velocity monitoring program of binary stars with evolved components. The program aims to obtain a clear picture of how these binary systems evolve \citep{vanwinckel10}. The spectroscopic data is obtained using the HERMES spectrograph mounted on the $1.2\,$m MERCATOR telescope in the Roque de los Muchachos observatory, Canary Islands, Spain \citep{raskin11}. The spectra have a spectral resolution $R=85\,000$ and cover a wavelength range from $3770$ to $9000\,$\angstrom. The total number of high-resolution optical spectra that we obtained ($N_\text{obs}$) and their average signal-to-noise in \halpha\,are listed in Table~\ref{tab:specdata} for each object. To optimise the fitting routine, the spectra should provide a good phase coverage of the orbital period, have a high signal-to-noise in \halpha\, (S/N$>20$), and be observed within four orbital periods of one another. Additionally, if an orbital phase\footnote{We define orbital phase as the phase (from 0 to 1) in the orbital period, where orbital phase 0 corresponds to periastron.} is over-represented by spectra, we remove the spectra with the lowest signal-to-noise for this phase, in order to avoid over-fitting. This leaves us with between 26 and 42 spectra for our program stars ($N_\text{used}$ in Table~\ref{tab:specdata}). In Figure~\ref{fig:dynspec}, we display the time-series of the \halpha\, profiles folded on the orbital phase. We call this representation the dynamic spectra and colour code the normalised spectrum per orbital phase. To make the dynamic spectra smooth, we interpolate in orbital phase. The total number of selected spectra are indicated with a horizontal dash on these dynamic spectra. This shows that we have a good coverage of the orbital phase for these program stars.


\subsection{Sample}\label{ssec:sample}

\begin{table} 
    \begin{center}
    \caption{Stellar parameters for the five objects in our sample. The stellar parameters are effective temperature $T_\text{eff}$, surface gravity $\log\,g$, metallicity [Fe/H], and luminosity of the post-AGB star $L_1$. a:~\citet{kipper11}; b: This work; c: \citet{gonzalez97a}; d:\citet{oomen20}; e: \citet{hrivnak08}; f: \citet{giridhar05}; g: \citet{manick17}; h: \citet{giridhar00}.}
    \label{tab:stellarpar}
    \begin{tabular}{l c c c c c}
        \hline
        Object     & $T_\text{eff}$ & $\log\,g$ & [Fe/H] & $L_1$     & Ref. \\
                   & K              &           & dex    & L$_\odot$ & \\
        \hline
        89\,Her    & 6600           & 0.8       & -0.5   & 9930          & a,b    \\
        EP\,Lyr    & 6200           & 1.5       & -1.8   & 5150          & c,d    \\
        HD\,46703  & 6250           & 1.0       & -1.5   & 2450          & d,e    \\
        HP\,Lyr    & 6300           & 1.0       & -1.0   & 8000          & a,f    \\
        TW\,Cam    & 4800           & 0.0       & -0.5   & 3000          & g,h    \\
        \hline
    \end{tabular}
    \end{center}
\end{table}

We aim to model a diverse set of post-AGB binaries. To that end, we select five objects based on three criteria: (1) The objects exhibit distinct temporal variability in their \halpha\,line caused by the jet. (2) The objects span a large range of orbital parameters. (3) Since we have already tested the spatio-kinematic model on \bd\, and \iras\, in \citet{bollen19} and \cite{bollen20}, we choose objects that differ from these two post-AGB binaries. Based on these criteria, we selected the following five jet-creating post-AGB binary systems for our analysis: 89\,Her, EP Lyr, HD 46703, HP Lyr, and TW Cam. We will introduce these objects in the following subsections.


\subsubsection{89\,Her}\label{sssec:89her}

89\,Her is an extensively studied post-AGB binary system. The binary nature of this system was first established by \citet{arellano84} and later confirmed by \citet{waters93}. \citet{bujarrabal07} examined CO interferometric data of 89\,Her and found the presence of two nebular components: a compact unresolved component from the circumbinary disc and an extended hour-glass structure with an expansion velocity of about $7\,$\kms, which can be ascribed to a molecular outflow. \citet{bujarrabal07} inferred a binary inclination angle of $15\degr$, based on the symmetry axis of the hour-glass outflow. \citet{hillen13, hillen14} established that the presence of a jet-like outflow originating from the companion can explain the presence of the over-resolved scattering component observed in the reconstructed image based on near-IR interferometric data and the P Cygni-like \halpha\,line of 89\,Her. \cite{gangi20} acquired spectra over a time-interval of 40 years and found long-term variability in the \halpha\,line. They propose a possible scenario of recursive mass-ejection events, occurring every $5000\,$d.

\cite{kipper11} analysed the chemical composition of 89\,Her and found the primary star to be metal deficient with [Fe/H] $= -0.5$. They also found $T_\text{eff} = 6600\,$K and $\log\, g = 0.8$ (see Table~\ref{tab:stellarpar}). With an orbital period of $289.1\,$days, 89\,Her has the shortest period of the objects in our program stars \citep{oomen18}. This binary has a high eccentricity of $0.29$. We use 26 high-resolution optical spectra of this object, which almost cover four full binary orbits. These spectra have high signal-to-noise ratios that range between $80$ and $160$ in the \halpha\,line.

The \halpha\,line of 89\,Her has a P Cygni-like profile at all orbital phases, where the absorption feature reaches blue-shifted velocities between $100-200\,$\kms. Interestingly, when we look at the line variations in the dynamic spectra of the \halpha\,line in Figure~\ref{fig:89_dynspec}, it becomes apparent that the orbital-phase dependent variations are minor. The absorption feature is present throughout the whole orbital phase and is at its strongest during superior conjunction, at orbital phase $\phi=0.45$. Since the absorption feature is always observed and the inclination angle of the binary system is rather small, it is highly likely that the continuum light from the post-AGB star towards the observer is blocked by the jet during the whole orbital period. This implies that the jet has an half-opening angle larger than the inclination angle of the binary system.


\subsubsection{EP Lyr}\label{sssec:eplyr}

EP\,Lyr is one of three RV Tauri stars in the sample. EP Lyr was classified as an RVa photometric variable by \citet{zsoldos95}, who identified pulsation periods of $41.6$ and $83.8\,$days. \citet{manick17} found similar pulsation periods, based on the radial velocity measurements of their spectroscopic data. Abundance analyses by \citet{gonzalez97a} indicated stellar parameters of $\teff=6200\,$K, $\log\,g=1.5$, and [Fe/H]$=-1.8$. However, due to the pulsations, the post-AGB star in the system shows significant variations in its effective temperature, which can fluctuate by $800\,$K \citep{gonzalez97a}. The binary system has a long period of $1151\,$days and the highest eccentricity ($e=0.39$) in our sample \citep{oomen18}. 

The 30 spectra that we use in this study cover two and a half orbital cycles and have a signal-to-noise ratio that averages at S/N$=45$ in \halpha. The main feature that stands out in the dynamic \halpha\,line of EP\,Lyr, shown in Figure~\ref{fig:ep_dynspec}, is the irregularity in the double-peaked emission. According to \citet{pollard97}, this sudden increased emission in the \halpha\,line can be attributed to shock events in the photosphere of the evolved star. The absorption feature, caused by the jet in EP\,Lyr, is short-lived but very prominent in the \halpha\,line. The jet blocks the light from the primary star during approximately $20\%$ of the orbital phase and reaches velocities up to $200\,$\kms.


\subsubsection{HD 46703}\label{sssec:hd46703}

HD\,46703 is a post-AGB binary system which shows mild flux variations in its photometric data, suggesting that the post-AGB star has weak pulsations \citep{hrivnak08}. \cite{hrivnak08} performed a chemical abundance analysis on HD\,46703 and found the following stellar parameters: effective temperature $\teff = 6250\,$K, surface gravity $\log\,g = 1.0$, and metallicity [Fe/H]$=-1.5$. HD\,46703 has orbital period of $597.4\,$days and an eccentricity of $0.30$ \citep{oomen18}. As mentioned by \citet{oomen20}, the inclination angle of the binary orbit is most likely smaller than $60$ degrees, since the object does not show an RVb signature.

For this object, we select 34 spectra that cover 2.7 orbital cycles. The signal-to-noise ratio of the spectra range between S/N$\,=27$ and S/N$\,=99$ in \halpha. The \halpha\,line of HD\,46703 is mainly governed by the photospheric absorption (See Figure~\ref{fig:hd_dynspec}). The line has a weak double-peaked emission feature, which follows the primary component, suggesting that this emission is associated to material in the vicinity of the primary component. The blue-shifted absorption feature caused by the jet is short-lived during the orbital period and relatively weak. This feature is not centred on the phase of superior conjunction. The absorption feature occurs sooner, suggesting that the jet might be tilted forward, towards the direction of travel of the companion.

\subsubsection{HP Lyr}\label{sssec:hplyr}

HP\,Lyr is the second RV Tauri object in our sample. This object was classified as a photometric RVa type by \citet{graczyk02}. They found a remarkably long pulsation period of $68.95\,$days. The spectroscopic analysis by \cite{manick17} found a similar period of $68\,$days from a time-series analysis of their spectroscopic data. Out of the 68 spectra obtained by \citet{manick17}, 17 show observational shock signatures, which indicates that shocks occur rather frequently in this object. The stellar parameters of HP\,Lyr were determined during the abundance analysis of \citep{giridhar05}: $\teff = 6300\,$K, $\log\,g = -1.0$, and [Fe/H]$=1.0$. HP\,Lyr has an orbital period of $1818\,$days and thus has the longest period in our sample. The eccentricity for this binary system is $0.20$.

We select 40 spectra of HP\,Lyr that cover $92\%$ of the full binary orbit, with an average signal-to-noise ratio of 53. Due to the long orbital period of this system, we did not obtain any observations between orbital phase $\phi=0.68$ and $\phi=0.77$. HP\,Lyr has one of the most remarkable jet absorption features in its \halpha\,line. The dynamic \halpha\,line is represented in Figure~\ref{fig:hp_dynspec}. The jet absorption is observed during more than half of the orbital period. The absorption feature rapidly increases in strength around orbital phase $0.2$, and appears to slowly drop of in strength, until it disappears at orbital phase $0.75$. During these phases, the maximum, blue-shifted extent of the absorption stays relatively constant at $-150\,$\kms. A weak emission is observed throughout the whole orbital phase, which occasionally shows peaks, caused by the shocks. Interestingly, although shocks are more prevalent in this system than in EP Lyr, the increased strength of the emission in \halpha\,is limited.


\subsubsection{TW Cam}\label{sssec:twcam}

TW\,Cam is the third RV Tauri star in our sample, with two pulsation periods of 43 and 85 days \citep{preston63, zsoldos91, manick17}. This object was first classified as a photometric type RVa star by \cite{giridhar00}, until \cite{manick17} showed long-term photometric variations in its light curve and reclassified TW\,Cam as a photometric type RVb. \cite{giridhar00} derived the following stellar parameters in their abundance analysis of TW\,Cam: $\teff = 4800\,$K, $\log\,g = 0.0$, and [Fe/H]$=-0.5$. This object has an orbital period of 662 days and an eccentricity of 0.25.

TW\,Cam has a very prominent jet absorption feature in its \halpha\,line, that covers about $50\%$ of the orbital period. The absorption shows peculiar variations during the orbital phase, as it peaks to a blue-shifted velocity of $-400\,$\kms\,twice during orbital phases 0.55 and 0.75, but drops to $-200\,$\kms\,between these phases, giving it a double-peaked shape. The emission feature is present throughout the whole orbital phase. Although \citet{manick17} observed shock features in the spectroscopic data of TW\,Cam, the shock-related emission peaks in the \halpha\,line are weak in strength.

\section{Methods}\label{sec:methods}

In this work, we apply the spatio-kinematic model and radiative transfer model from \cite{bollen20}, in order to determine the geometric, kinematic, and density structure of the jet and its mass-loss rate. Below, we briefly describe the two modelling techniques. We refer the reader to \cite{bollen19} and \cite{bollen20} for a detailed description.

For the spatio-kinematic modelling of the jet, we recreate the observed absorption feature in \halpha\,caused by the scattering of photospheric light of the post-AGB star, by modelling the jet-binary system. The main components of the model are the post-AGB star, the companion star, and the jet, which is centred on the companion. 

In the model, we trace the light that travels from the post-AGB star towards the observer. Hence, we first provide the model with a background spectrum, which does not include the jet absorption feature. This background spectrum has two components: the photospheric light from the post-AGB star and an emission component observed in \halpha. For all five program stars, the emission component in \halpha\, follows the radial velocity motion of the post-AGB star. Hence, we assume that this emission is linked to the post-AGB star. 

If this (background) light passes through the jet, we calculate the optical depth along this ray in order to determine how much light is scattered by the jet. By doing so, we create synthetic \halpha\, spectra for each orbital phase. We then fit the synthetic \halpha\,line variations to the observed \halpha\,lines using the \texttt{emcee} Markov chain Monte Carlo (MCMC) algorithm \citep{foreman13}. This gives us the best-fitting parameters for the jet-binary model. 

In the model, we implement three jet configurations: the stellar jet, the X-wind, and the disc wind. The stellar jet is based on the jet model by \cite{matt05}, where part of the disc matter is ejected along the open stellar magnetic field lines that are anchored on the star. The disc wind configuration is based on the magneto-centrifugal disc wind model by \cite{blandford82}. In this model, the matter is ejected along the magnetic field lines anchored to the accretion disc. These magnetic field lines have a minimum inclination of $30\,$\degr\, with respect to the jet axis. The X-wind model by \cite{shu94} is similar in nature to the disc wind model. The disc matter is also launched at an angle larger than $30\,$\degr. However, the magnetic field in this model is squeezed together in the inner disc region, giving the outflow its distinct X-shape. We incorporate the most important features of these three jet models in our jet configurations. Just as the model by \cite{matt05}, the stellar jet is centred on the companion and the jet matter follows the magnetic field lines of the star. For the X-wind configuration, we include a jet cavity which has a minimum angle of $25\,$\degr. The disc wind configuration is similar to the X-wind, but the ejected matter in the model originates from the surface of the circum-companion accretion disc. Additionally, we allow the jet in each configuration to be inclined with respect to the orbital axis of the binary system.

Next, we use the best-fitting jet configuration of the spatio-kinematic model as input for the radiative transfer modelling of the jet. For the radiative transfer, we assume local thermodynamic equilibrium and an isothermal jet. This radiative transfer model calculates the level populations of the excited states of hydrogen, in order to calculate the amount of scattering that is caused by the jet in the Balmer lines. These level populations are dependent on the local temperature and density of the jet. Hence, the total amount of absorption in the four Balmer lines and their relative difference will change for different jet temperatures and densities. In order to quantify this absorption, we calculate the equivalent width (EW) of the first four Balmer lines in the observations, i.e., \halpha, \hbeta, \hgamma, and \hdelta. Then, we use a grid of jet temperatures and densities to create synthetic Balmer line spectra and fit the EW of these synthetic lines to the EW of the observed spectra. By doing so, we can find the jet densities and temperatures, from which we can determine the jet mass-loss rates in each system.


\section{Results}\label{sec:results}

\begin{figure*}
  \centering
    \begin{subfigure}[b]{.5\linewidth}
        \centering\large 
          \includegraphics[width=1\linewidth]{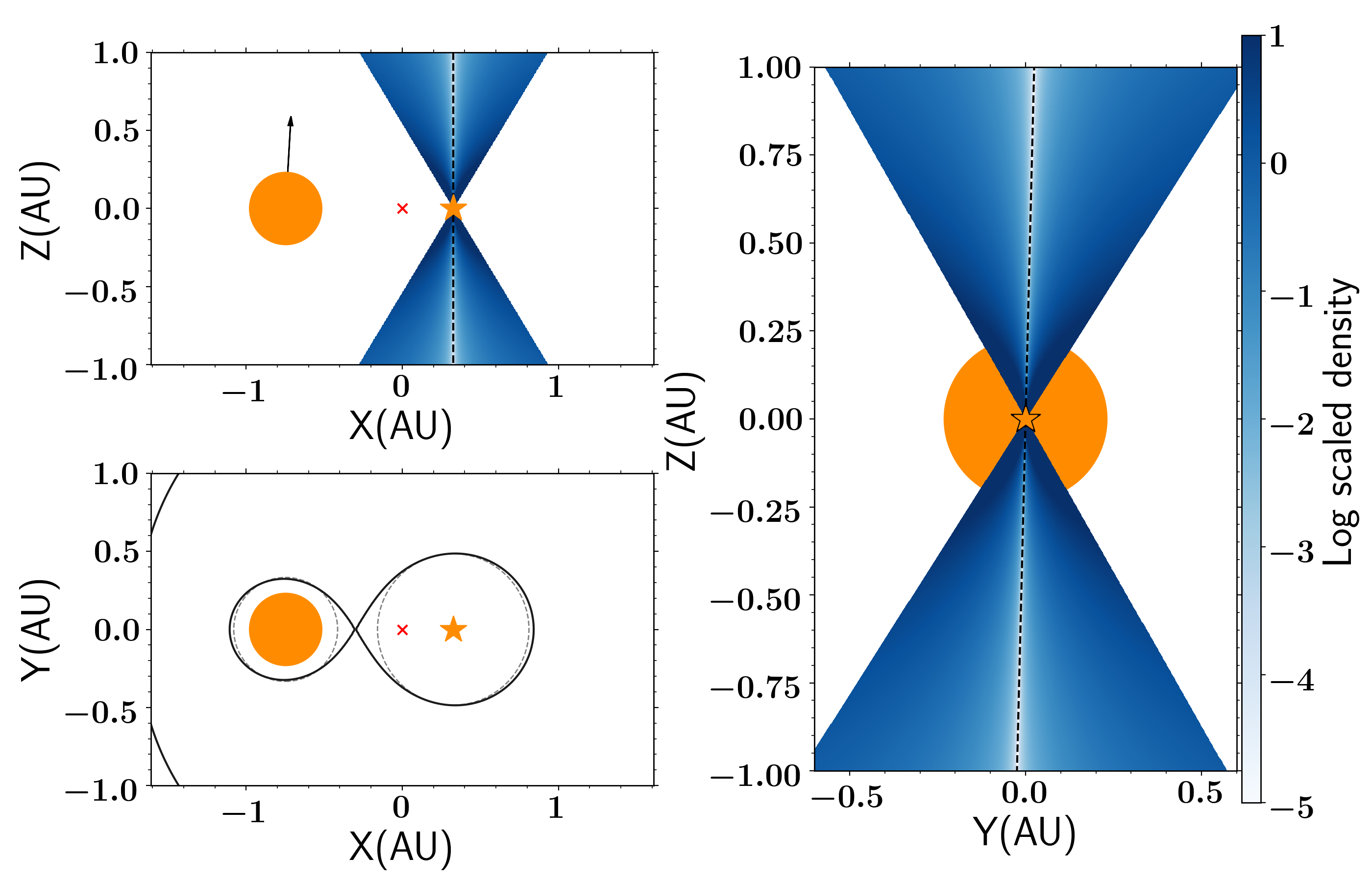}
        \caption{89\,Her}\label{fig:geom_89her}
    \end{subfigure}%
    \begin{subfigure}[b]{.5\linewidth}
        \centering\large 
          \includegraphics[width=1\linewidth]{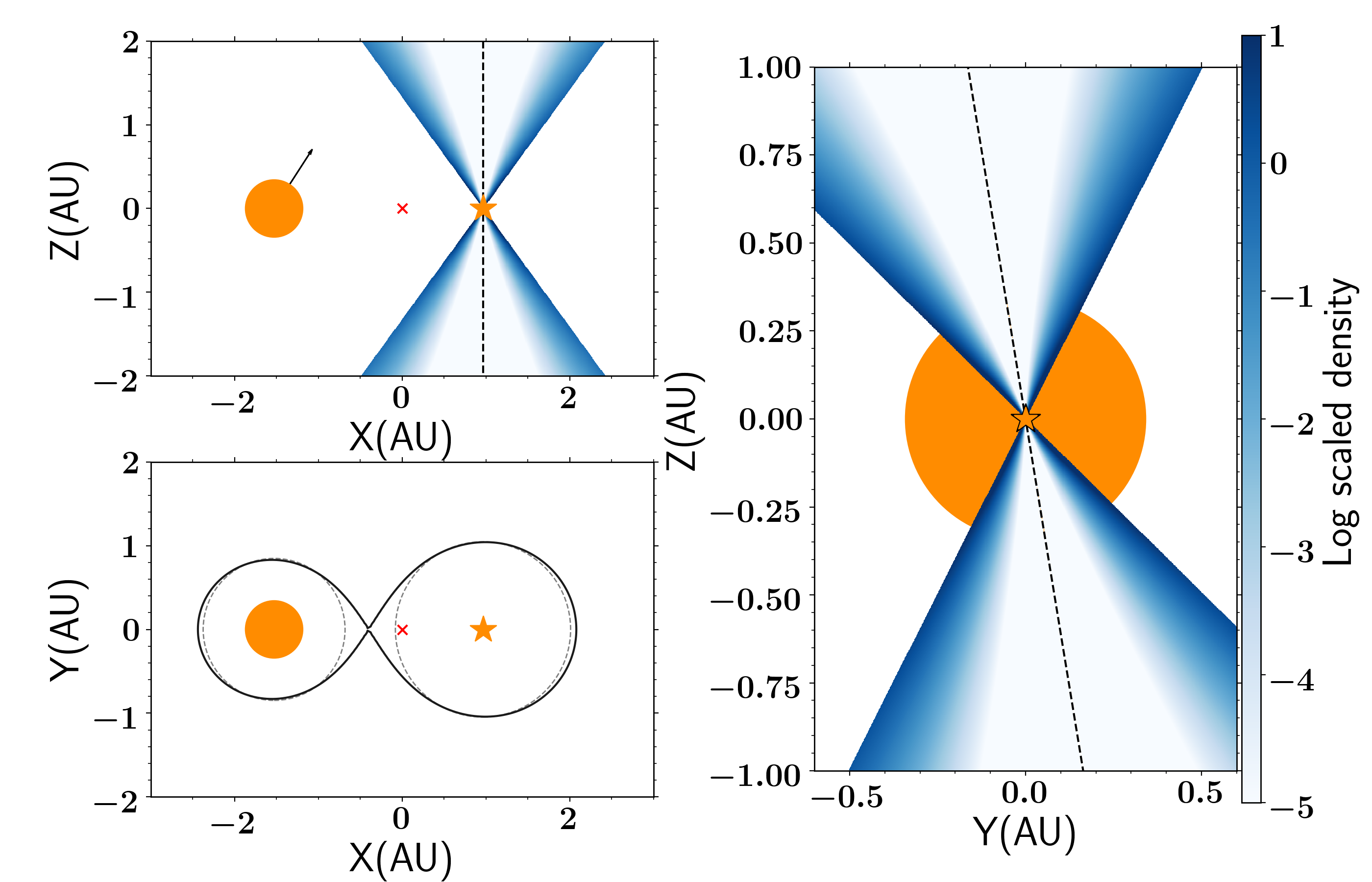}
        \caption{EP Lyr}\label{fig:geom_eplyr}
    \end{subfigure}%
    
    \begin{subfigure}[b]{.5\linewidth}
        \centering\large 
          \includegraphics[width=1\linewidth]{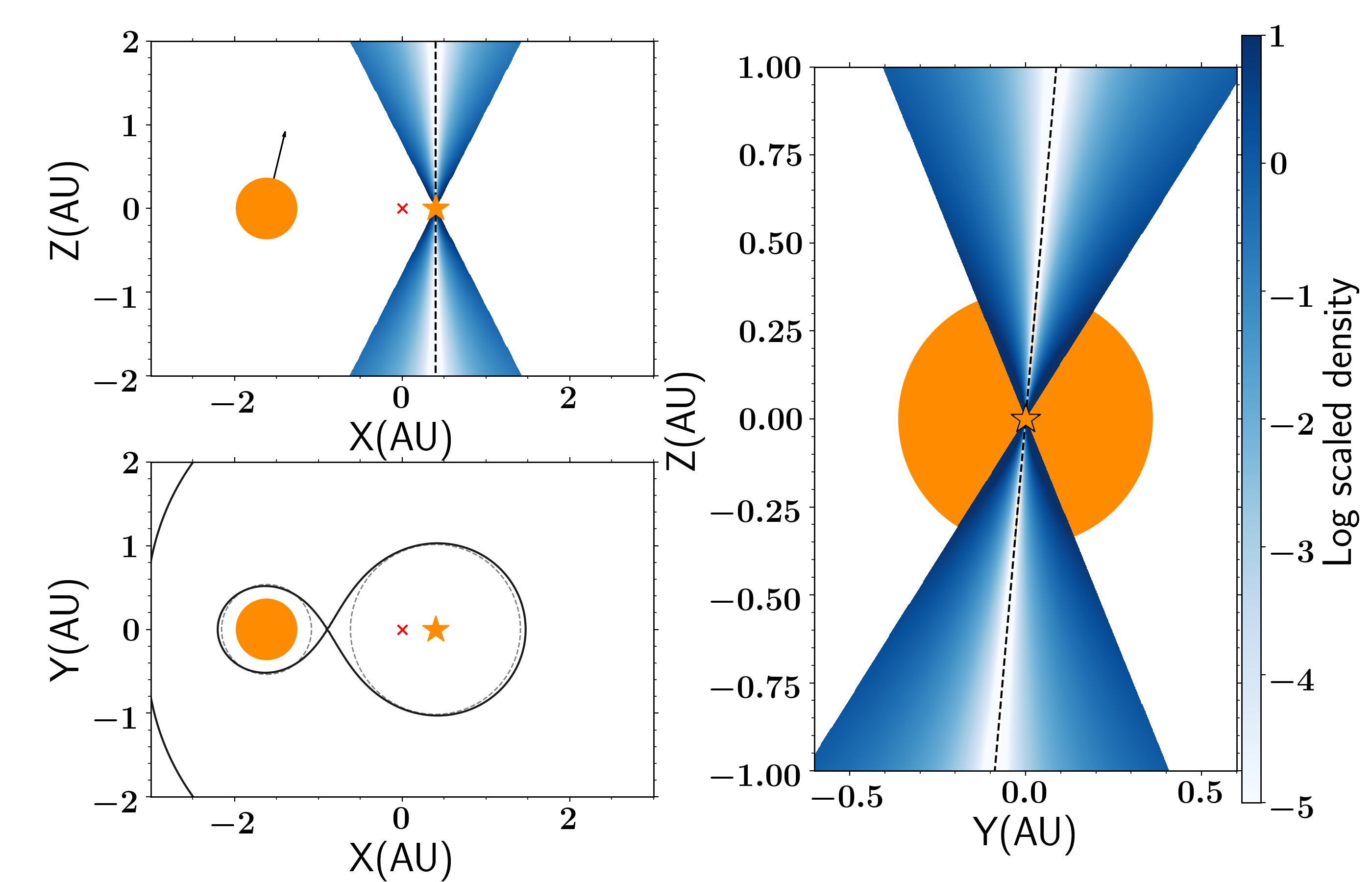}
        \caption{HD 46703}\label{fig:geom_hd}
    \end{subfigure}%
    \begin{subfigure}[b]{.5\linewidth}
        \centering\large 
          \includegraphics[width=1\linewidth]{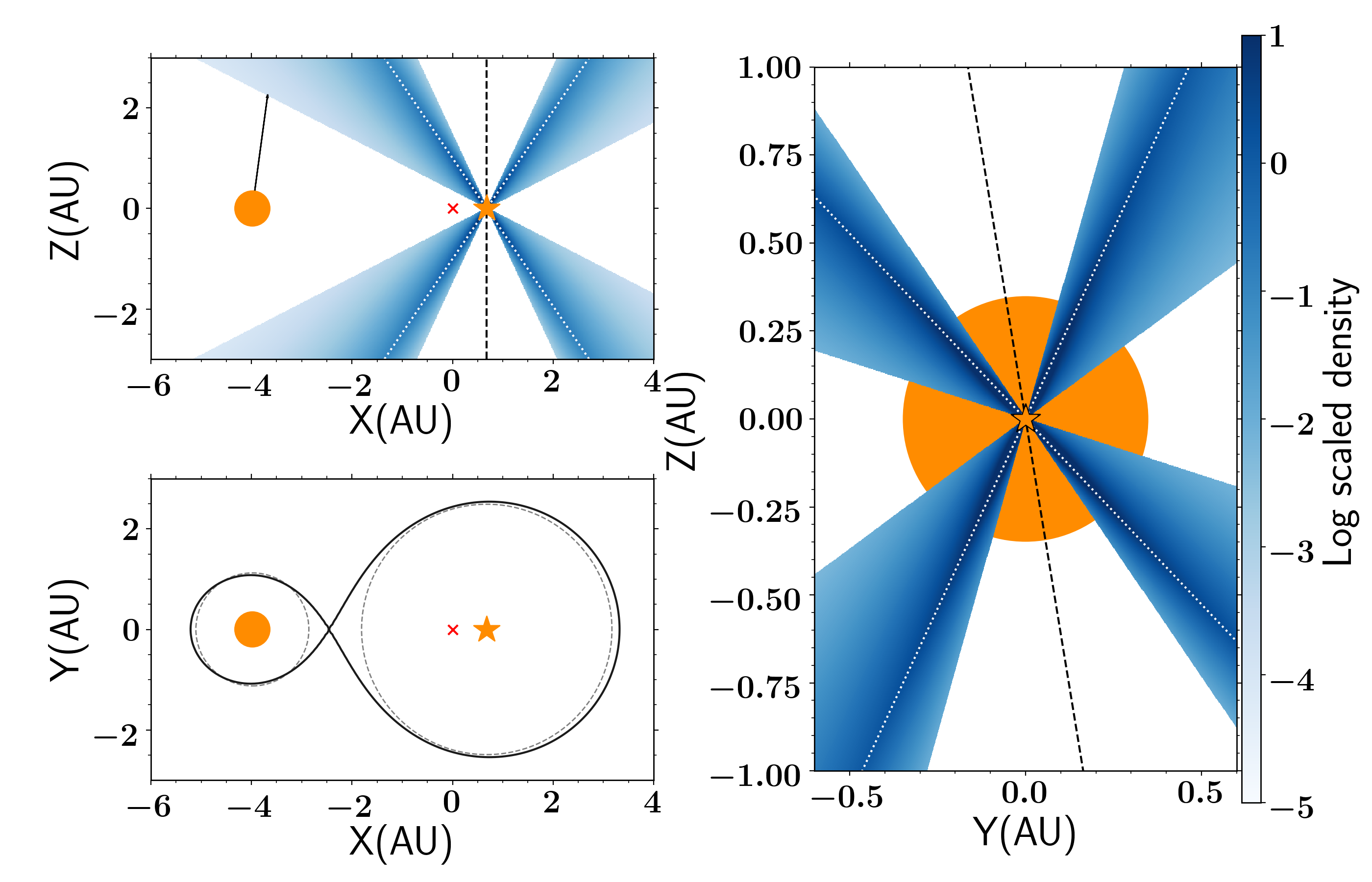}
        \caption{HP Lyr}\label{fig:geom_hplyr}
    \end{subfigure}%
    
    \begin{subfigure}[b]{.5\linewidth}
        \centering\large 
          \includegraphics[width=1\linewidth]{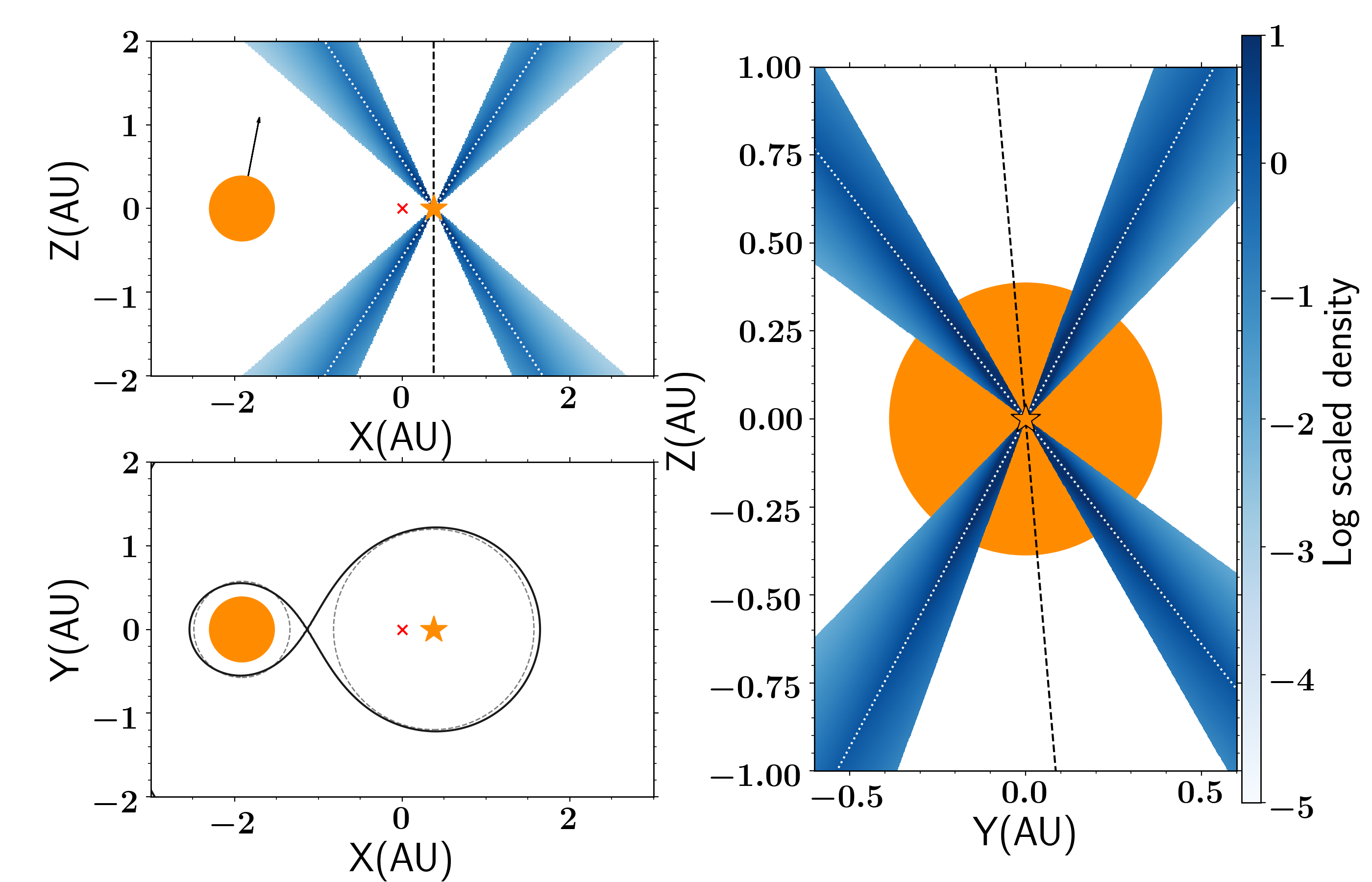}
        \caption{TW Cam}\label{fig:geom_twcam}
    \end{subfigure}%
    \caption{The geometry of the binary system and the jet for the five objects, based on the results of the spatio-kinematic modelling. The plots show the system during superior conjunction, when the post-AGB star is directly behind the companion, as viewed by the observer. In each plot, the full orange circle represents the post-AGB star. The orange star symbol indicates the location of the companion. The red cross is the location of the centre of mass of the binary system. The radius of the post-AGB star is to scale. The jet is represented in blue, where the colour indicates the relative density of the jet. The dashed black line is the jet axis and the dotted white lines are the inner jet edges. The jet cavity is the inner region of the jet. The plot in the upper-left panel of each object shows the system viewed along the orbital plane, from a direction perpendicular to the line-of-sight to the observer. The right-hand-side panel shows the system viewed from a direction perpendicular to the X-axis. The post-AGB star is located behind the companion and its jet in this image. The lower-left panel shows the binary system viewed from above, perpendicular to the orbital plane. The grey dashed lines represent the Roche radii of the two binary components and the full black line shows the Roche lobes. } \label{fig:geom}
    
\end{figure*}

\begin{table*} 
    \begin{center}
    \caption{Best-fitting jet configuration, and jet parameters for the spatio-kinematic model of 89\,Her, EP Lyr (excluding and including pulsations and shocks), HD 46703, HP Lyr, and TW Cam. For comparison purposes, we also include the reduced chi-square values for the three jet configurations. The tabulated parameters are: inclination angle of the binary system $i$, jet outer angle $\theta_\text{out}$, jet inner angle $\theta_\text{in}$, jet cavity angle  $\theta_\text{cav}$, jet tilt $\phi_\text{tilt}$ , inner jet velocity  $v_\text{in}$, jet velocity at the jet edges  $v_\text{out}$, exponent for the velocity profile  $p_\text{v}$, exponent for the density profile for the outer and inner region $p_{\rho \text{,out}}$ and  $p_{\rho\text{,in}}$, optical depth scaling parameter  $c_\tau$, the radius of the post-AGB star $R_\text{1}$, and the reduced chi-square for the three jet configurations. The objects with a stellar jet configuration only have an outer jet angle $\theta_\text{out}$ and one exponent for the density profile  $\theta_\text{out}$. The errors are the 1$\sigma$ uncertainty interval and are statistical only.}
    \label{tab:bestfit}
    \begin{tabular}{l c c c c c c}
        \hline
                  & 89\,Her & EP Lyr & EP Lyr - puls & HD 46703 & HP Lyr & TW Cam \\
        Parameter &         &        &               &          &  
                  & \\
        \hline
        Best-fit configuration                       &   stellar jet     &  stellar jet      &  stellar jet     &  stellar jet     &  X-wind           &  X-wind          \\
         $i$ ($\degr$)                      &   $8.2\pm0.2$     &  $58.2\pm0.3$     &  $58.0\pm0.3$    &  $31.2\pm0.3$    &  $18.6\pm0.7$     &  $25.7\pm0.2$    \\
         $\theta_\text{out}$ ($\degr$)      &   $31.1\pm0.1$    &  $36.0\pm0.3$     &  $35.4\pm0.4$    &  $27.1\pm0.3$    &  $62.8\pm0.8$     &  $48.8\pm0.8$    \\
         $\theta_\text{in}$ ($\degr$)       &                   &                   &                  &                  &  $34.2\pm0.5$     &  $33.1\pm0.3$    \\
         $\theta_\text{cav}$ ($\degr$)      &                   &                   &                  &                  &  $25.0\pm0.3$     &  $25.0\pm0.1$    \\
         $\phi_\text{tilt}$ ($\degr$)       &   $-1.4\pm0.3$    &  $9.3\pm0.2$      &  $9.0\pm0.3$     &  $-5.00\pm0.01$  &  $9.6\pm0.3$      &  $4.9\pm0.1$     \\
         $v_\text{in}$ (km\,s$^{-1}$)       &   $155\pm1$       &  $280\pm5$        &  $300\pm9$       &  $177\pm6$       &  $211\pm3$        &  $443\pm4$       \\
         $v_\text{out}$ (km\,s$^{-1}$)      &   $0.15\pm0.10$   &  $32.3\pm1.3$     &  $32\pm2$        &  $0.11\pm0.02$   &  $11\pm2$         &  $40\pm3$        \\
         $p_\text{v}$                       &   $-0.41\pm0.06$  &  $-2.9\pm0.2$     &  $-2.5\pm0.2$    &  $0.5\pm0.1$     &  $8.5\pm0.2$      &  $9.9\pm0.1$     \\
         $p_{\rho\text{,in}}$               &                   &                   &                  &                  &  $14.9\pm0.2$     &  $8.1\pm0.3$     \\
         $p_{\rho\text{,out}}$              &   $1.88\pm0.07$   &  $14.8\pm0.4$     &  $14.9\pm0.4$    &  $4.3\pm0.2$     &  $-12.1\pm0.2$    &  $-14.5\pm0.3$   \\
         $c_\tau$                           &   $2.67\pm0.01$   &  $2.82\pm0.03$    &  $2.89\pm0.02$   &  $2.05\pm0.02$   &  $3.08\pm0.03$    &  $2.97\pm0.002$  \\
         $R_\text{1}$ (AU)                  &   $0.231\pm0.001$ &  $0.341\pm0.005$  &  $0.32\pm0.01$   &  $0.36\pm0.01$   &  $0.347\pm0.005$  &  $0.386\pm0.003$ \\
         \hline
         $\chi^2_{\nu,\text{stellar}}$      &   $0.3$          &  $1.7$            &  $1.7$          &  $1.8$           &  $3.3$            &  $1.6$          \\
         $\chi^2_{\nu,\text{X-wind}}$       &   $0.7$          &  $3.4$            &  $8.2$           &  $3.2$           &  $1.7$            &  $1.3$          \\
         $\chi^2_{\nu,\text{disc-wind}}$    &   $0.7$          &  $2.1$            &  $2.2$           &  $2.0$           &  $2.4$            &  $1.7$          \\
        \hline
    \end{tabular}
    \end{center}
\end{table*}
\begin{figure*}
  \centering
    \begin{subfigure}[b]{.5\linewidth}
        \centering\large 
          \includegraphics[width=1\linewidth]{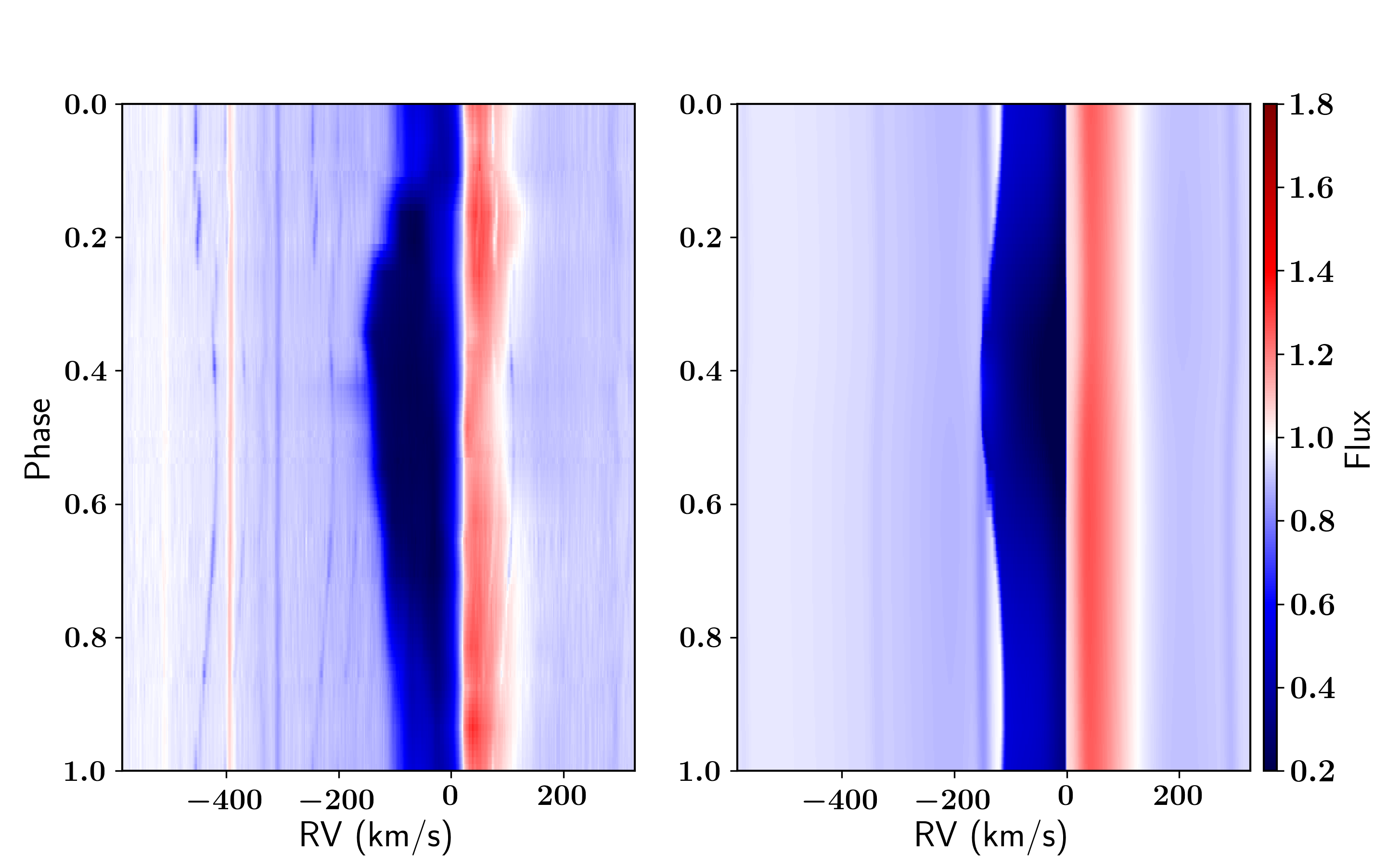}
        \caption{89\,Her}\label{fig:obsmod_89her}
    \end{subfigure}%
    \begin{subfigure}[b]{.5\linewidth}
        \centering\large 
          \includegraphics[width=1\linewidth]{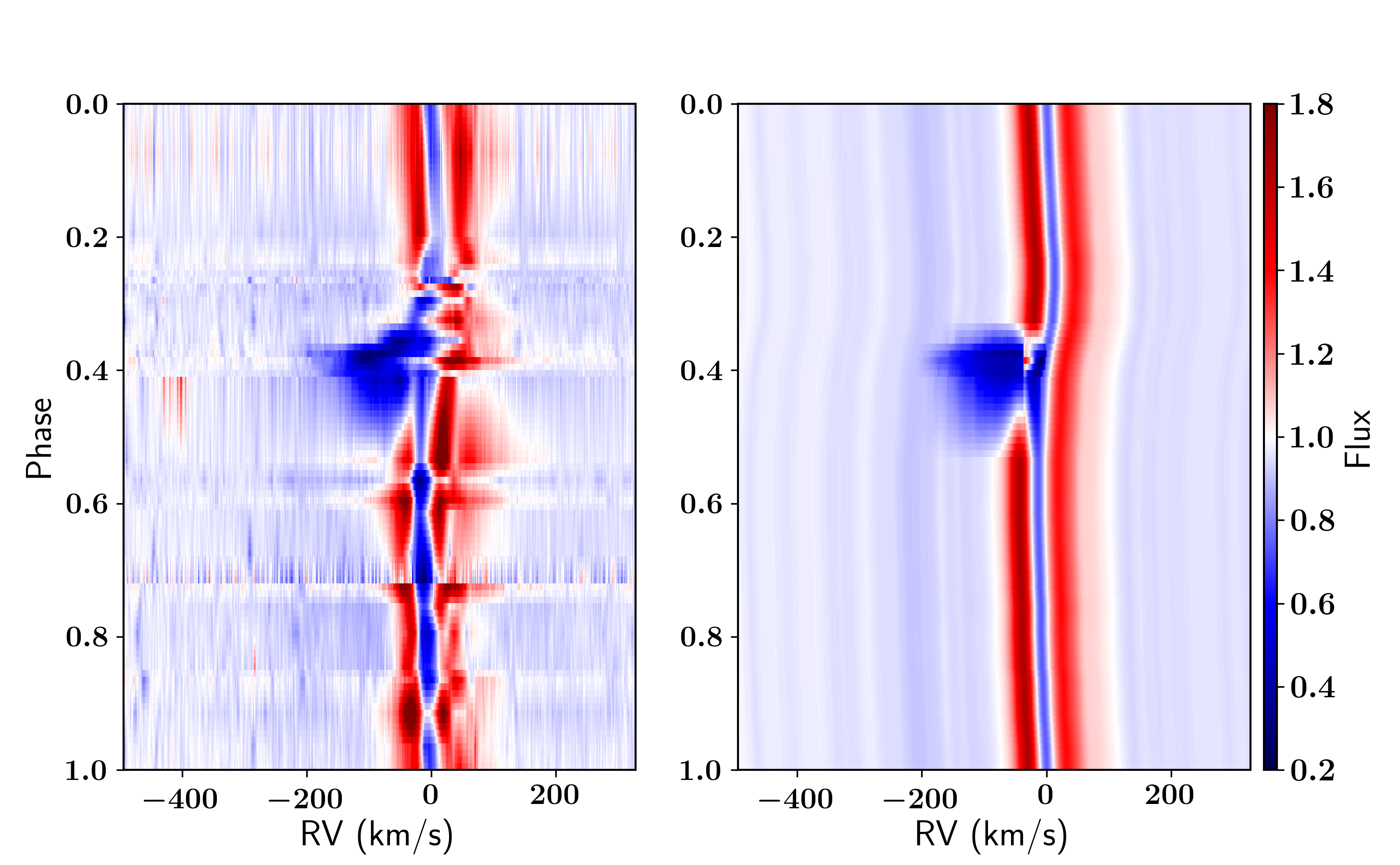}
        \caption{EP Lyr}\label{fig:obsmod_eplyr}
    \end{subfigure}%

    \begin{subfigure}[b]{.5\linewidth}
        \centering\large 
          \includegraphics[width=1\linewidth]{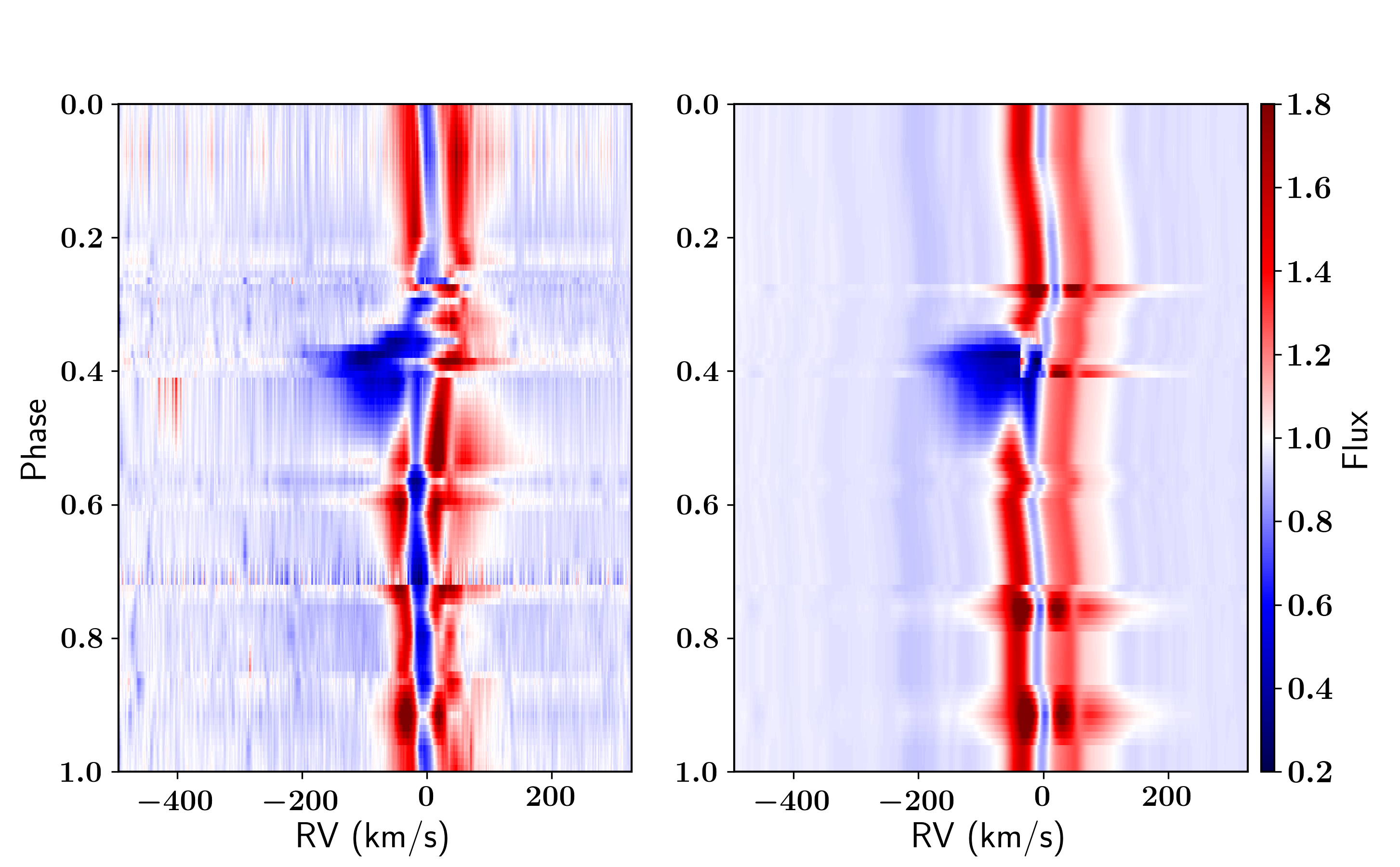}
        \caption{EP Lyr - including pulsations and shocks}\label{fig:obsmod_eplyr_shock}
    \end{subfigure}%
    \begin{subfigure}[b]{.5\linewidth}
        \centering\large 
          \includegraphics[width=1\linewidth]{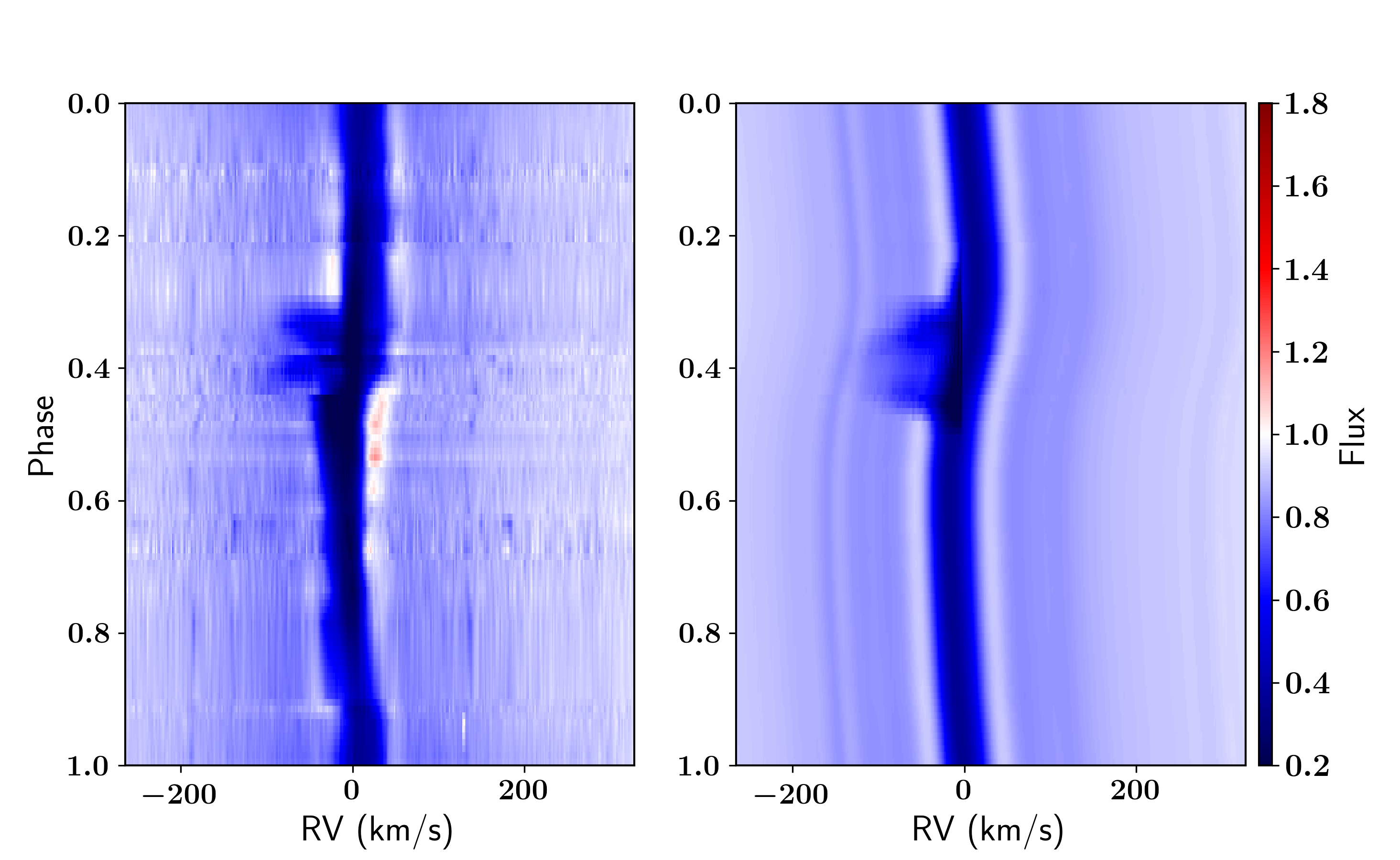}
        \caption{HD 46703}\label{fig:obsmod_hd}
    \end{subfigure}%
    
    \begin{subfigure}[b]{.5\linewidth}
        \centering\large 
          \includegraphics[width=1\linewidth]{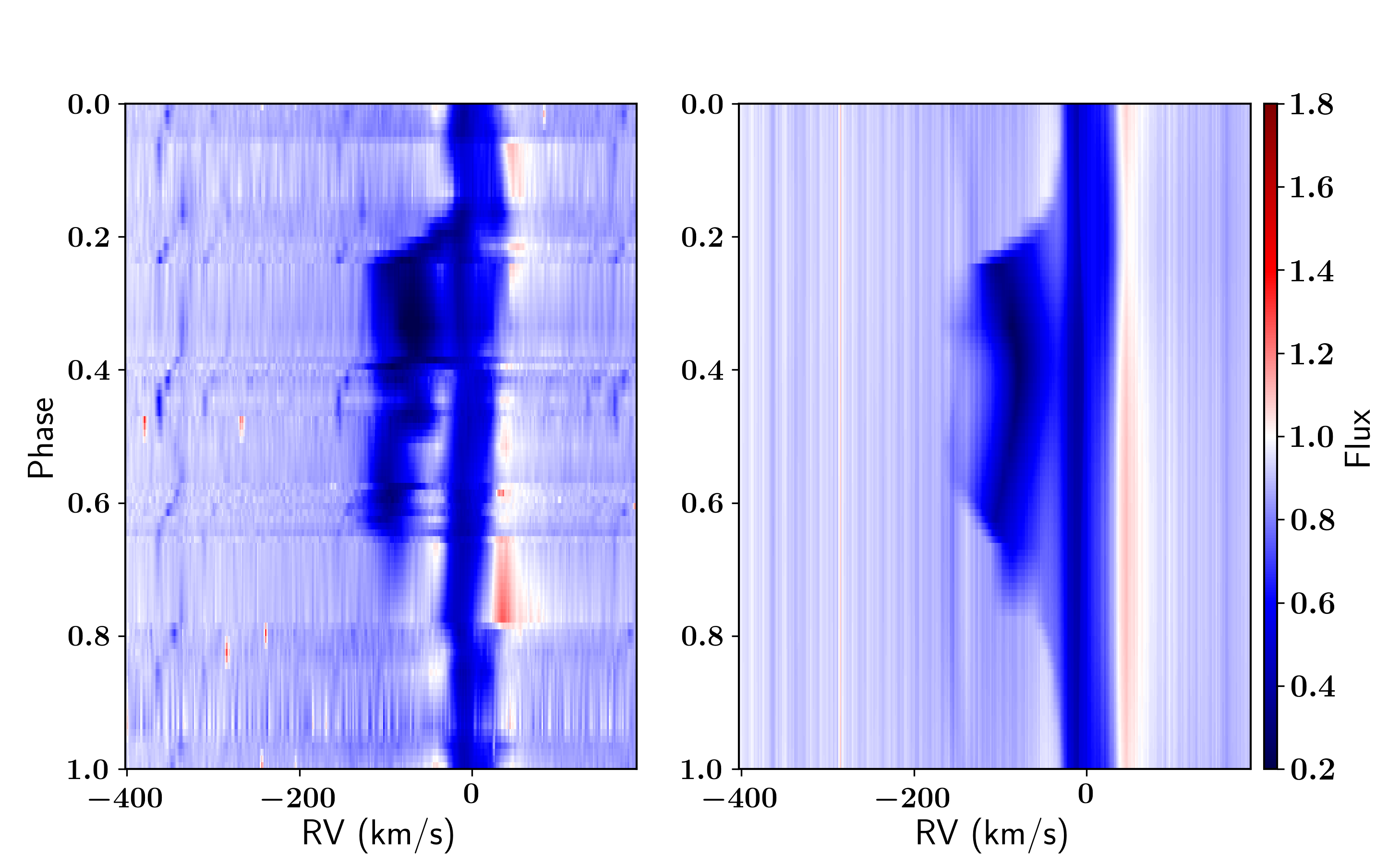}
        \caption{HP Lyr}\label{fig:obsmod_hplyr}
    \end{subfigure}%
    \begin{subfigure}[b]{.5\linewidth}
        \centering\large 
          \includegraphics[width=1\linewidth]{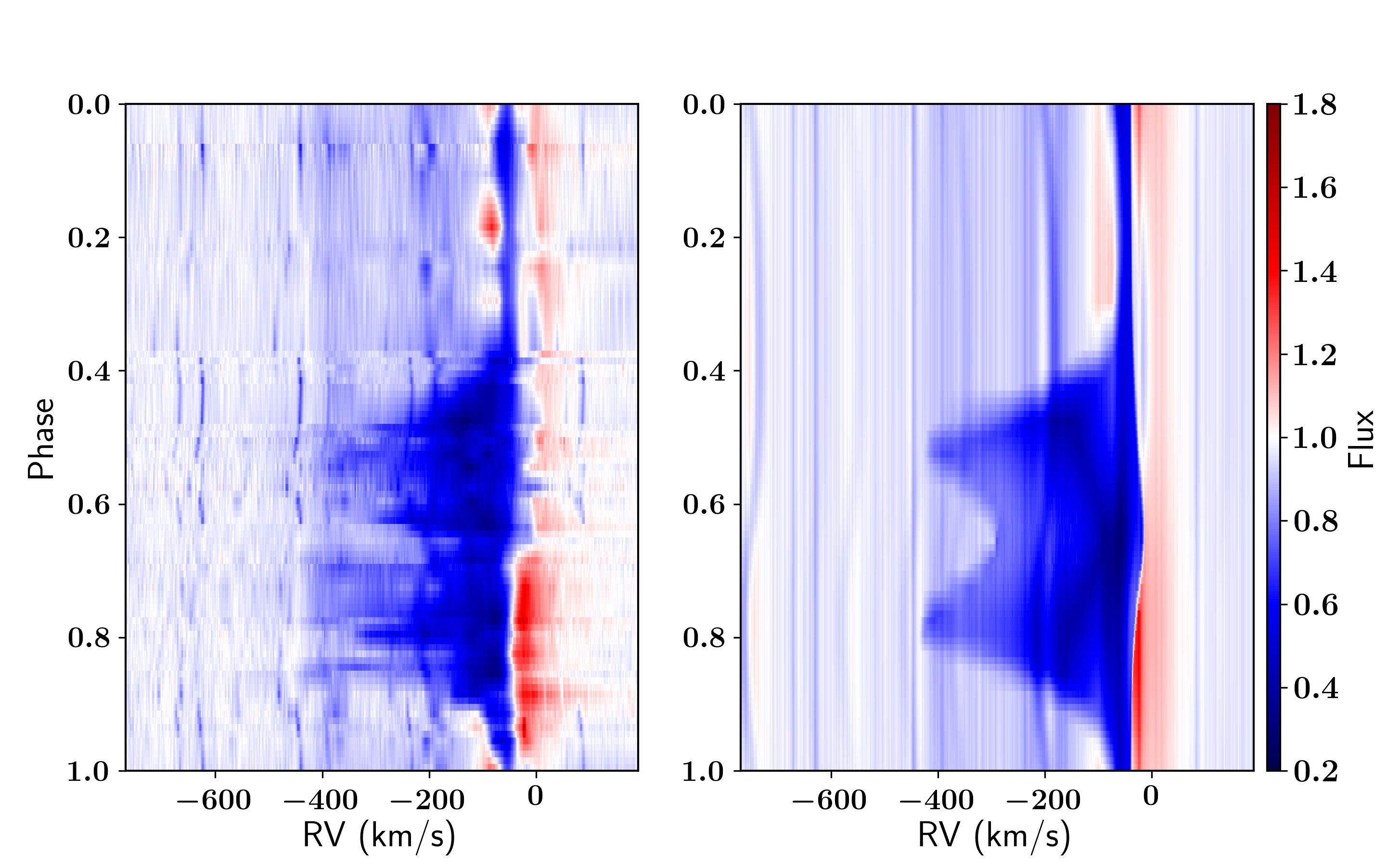}
        \caption{TW Cam}\label{fig:obsmod_twcam}
    \end{subfigure}%
    \caption{The dynamic spectra in \halpha\,for the observations (\textit{left}) and best-fitting models (\textit{right}) of the five objects in our sample. For EP Lyr, we show the best-fitting model for the fitting which excluded (\textit{panel b}) and included (\textit{panel c}) the pulsations and shocks. The spectra are plotted as a function of orbital phase. The velocity on the x-axis denotes the RV shift from the systemic velocity of the binary system. The spectra are normalised to the continuum-flux level.} \label{fig:dynspec_obsmod}
   
\end{figure*}

\begin{figure*}
\centering
\includegraphics[width=.88\textwidth]{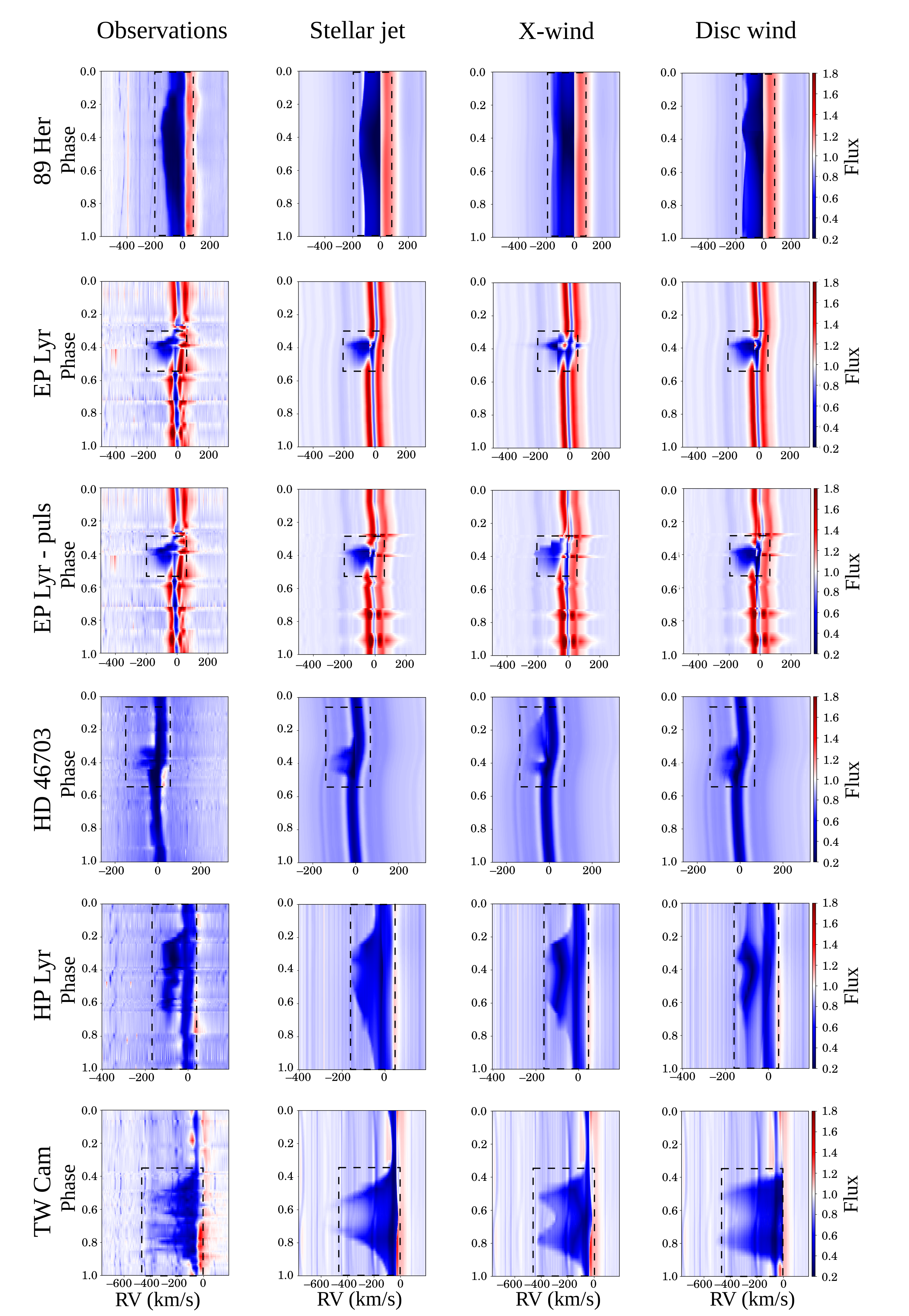}
\caption{The observed and modelled dynamic spectra for the best-fitting result of each object and each configuration. The dashed-box marks the region used to calculate the chi-square.}\label{fig:allmodels}
\end{figure*}

\begin{table} 
    \begin{center}
    \caption{Best-fitting jet temperature and density from the radiative transfer modelling. The given densities represent the density at the edge of the jet at a height of 1\,AU from the jet base.}
    \label{tab:bestTrho}
    \begin{tabular}{l c c}
        \hline
        Object     & $T_\text{jet}$ & $\log\,\rho$  \\
                   & K              &    kg/m$^3$       \\
        \hline
        89\,Her    & $5000^{+300}_{-300}$             & $17^{+0.6}_{-0.2}$       \\
        EP\,Lyr    & $5800^{+300}_{-1600}$            & $15.8^{+1.4}_{-1}$       \\
        HD\,46703  & $5400^{+200}_{-1300}$            & $16.1^{+0.8}_{-1.5}$       \\
        HP\,Lyr    & $5400^{+200}_{-900}$             & $16.6^{+0.9}_{-0.3}$      \\
        TW\,Cam    & $4400^{+200}_{-400}$             & $18.4^{+0.9}_{-0.2}$       \\
        \hline
    \end{tabular}
    \end{center}
\end{table}
In the following subsections, we describe the setup for the spatio-kinematic model for each object, followed by a description of the spatio-kinematic and radiative transfer results. The best-fitting jet configurations and parameters are listed in Table~\ref{tab:bestfit}. The best-fitting jet temperature and density are listed in Table~\ref{tab:bestTrho}. The corresponding geometry of the binary system and the jet are represented in Figure~\ref{fig:geom} and the synthetic spectra are shown in Figure~\ref{fig:dynspec_obsmod}. In order to compare the goodness-of-fit result between the three models, we calculate the reduced chi-square $\chi^2_\nu$ from the observed and modelled flux for each flux point in each spectrum.
\begin{equation}
    \chi^2_\nu = \frac{1}{N}\sum_i^{N_s}\sum_w^{N_\lambda}\frac{\big(F^o_{i,w} - F^m_{i,w}\big)^2}{\sigma_{i,w}^2}
\end{equation}
with $N=N_s \cdot N_\lambda$ the total number of data points, $N_s$ the number of spectra, $N_\lambda$ the number of wavelength bins, $F^o_{i,w}$ and $F^m_{i,w}$ the observed and modelled flux in wavelength bin $w$ of spectrum $i$, and $\sigma^2_{i,w}$ the standard deviation. The region that is used to calculate the chi-square for each model is indicated by the dashed box in Figure~\ref{fig:allmodels}.

The uncertainty for each parameter in Table~\ref{tab:bestfit} is determined from the posterior density distribution of the MCMC-modelling. The quoted uncertainties represent the 1$\sigma$\,interval. We discuss the chi-square values of the three configurations in Section~\ref{ssec:nature}. In Appendix~\ref{tab:allbestfit}, we present the best-fitting parameters for all three configurations. We note that our model is very sensitive to a change in parameters, which results in very small uncertainties. These uncertainties are not representative of the accuracy of the model parameters, however, but rather of the sensitivity of the parameters.


\subsection{Results: 89 Her}\label{ssect:results89her}

We perform the spatio-kinematic modelling of 89\,Her, using the three aforementioned jet configurations. We use a synthetic stellar spectrum from \citet{coelho14}, based on the atmospheric parameters of 89\,Her (see Section~\ref{sssec:89her}). We use the synthetic stellar spectra from this library for the other objects in our study as well. In the MCMC-modelling of 89\,Her, we limit the range for the inclination angle of the binary system between $8\degr$ and $20\degr$, based on the observed inclination angles found for 89\,Her \citep{bujarrabal07, hillen14}.

The best-fitting configuration for 89\,Her is a stellar jet, with a half-opening angle of $\theta = 31\degr$ and a jet velocity up to $v_\text{in} = 155\,$\kms. As can be seen in Figure~\ref{fig:geom_89her}, the density profile of the jet does not drop dramatically from the outer regions to the inner regions of the jet, but has a rather uniform density. With $-1\degr$, the jet shows no significant tilt with respect to the orbital axis of the binary\footnote{A positive jet tilt corresponds to a jet with the cone through which the line-of-sight passes through tilted away from the direction of travel of the companion. A negative tilt implies that this cone is tilted towards the direction of travel.}. The radius of the post-AGB star in 89\,Her is the smallest in our sample with $R_1=50\,$R$_\odot$.

The modelled dynamic spectra of 89\,Her reproduces the absorption feature well, which is observed over the whole orbital period of the binary system (see Figure~\ref{fig:obsmod_89her}). As we speculated in Section~\ref{sssec:89her}, this effect is due to an inclination angle that is smaller than the jet half-opening angle, such that our line-of-sight towards the primary star is always blocked by the jet. This is confirmed by our best-fitting model, where we find an inclination angle of $8\degr$ and a jet half-opening angle of $31\degr$ for 89\,Her.

We use this best-fitting jet configuration as input for the radiative transfer modelling. We fit for a range of jet densities and uniform jet temperatures. The jet temperatures range between $4400\,$K and $6600\,$K in steps of $100\,$K and the logarithm of the density at the edge of the jet and at a height of $1\,$AU from the jet base ranges between 14 and 18 in logarithmic steps of 0.2. We fit the EW in \halpha, \hbeta, \hgamma, and \hdelta\, from the model to the observations (see Figure~\ref{fig:ewfit_89}). The best fitting jet temperature and density are $T_\text{jet}= 5000^{+300}_{-300}\,$K and $\log\rho = 17^{+0.6}_{-0.2}$ (see Figure~\ref{fig:chisq_89}). This translates to a jet mass-loss rate of $2.0^{+7}_{-0.6}\times10^{-6}\,$\myr.


\subsection{Results: EP Lyr}\label{ssect:resultseplyr}
 
The observed \halpha\,line of EP\,Lyr shows strong variations due to the stellar pulsations and shocks of the primary. For this reason, we perform the MCMC-modelling for this object twice, with each run having different background spectra. For the first run, the background spectra include the photospheric light of the primary and the observed emission feature. This emission feature is always centred on the radial velocity of the primary and constant in strength. For the second run, we include these two components as well. We also include the radial velocity shift of the pulsations that were found by \citet{manick17} and the shocks in the \halpha\,line. These shocked spectra have a stronger emission component. The emission strength for these spectra is based on the average strength of the observed spectra with shocks. For both runs, we set a range of inclination angles between $30\degr$ and $80\degr$. 

The best-fitting model for the fitting of EP\,Lyr with and without pulsations is a stellar jet. Interestingly, the reduced chi-square of the model without pulsations and shocks is $\chi_\nu^2=1.7$, which is insignificantly better than the model with pulsations and shocks ($\chi_\nu^2=1.71$). This is mainly due to the irregularity of the occurrence and strength of the shocks, which makes it more complex to simulate the \halpha\,line. The results for EP Lyr with or without shocks are very similar and the values of the best-fitting parameters differ by less than $15\%$. For the remainder of our study, we will focus on the best-fitting model without pulsations.

The structure of the jet in EP\,Lyr is illustrated in Figure~\ref{fig:geom_eplyr}. The jet has a half-opening angle of $36\degr$ and is tilted by $9\degr$. The highest jet velocities reach $280\,$\kms. At its edges, the jet velocity has lower outflow velocities of $32\,$\kms. The density profile of the jet has a large exponent of $p_{\rho\text{,out}}=14.8$. This means that most of the mass will be located at the jet edges and the inner regions of the jet have an extremely low density, which is very similar to the jet cavity in the X-wind and disc wind configurations.

The modelled dynamic \halpha\,spectra of EP\,Lyr, shown in Figure~\ref{fig:obsmod_eplyr}, have a similar short-lived jet absorption feature as the observed dynamic spectra. Based on the best-fitting model results, this is induced by an inclination angle ($i=58\degr$) that is significantly larger than the jet half-opening angle ($\theta_\text{out}=36\degr$). In other words, this causes the line-of-sight from the observer to the primary star to be blocked by the jet for a brief period of time ($\sim20\%$ of the orbital period).

For the radiative transfer modelling, we set a range from $4200\,$K to $6200\,$K in steps of $100\,$K for the jet temperature and from 14 to 18 in steps of 0.2 for the logarithm of the jet density. For this object, we only fit the EW of \halpha\,and \hbeta. The best-fitting results of jet temperature and density are $T_\text{jet}= 5800^{+300}_{-1600}\,$K and $\log\rho = 15.8^{+1.4}_{-1}$ (see Figure~\ref{fig:chisq_eplyr}). The corresponding jet mass-loss rate for this jet density and configuration is $9\times10^{-8}\,$\myr. Due to the larger uncertainty in the data, the one-sigma confidence interval has a large range from $9\times10^{-9}\,$\myr\, to $2\times10^{-6}\,$\myr.


\subsection{Results: HD 46703}\label{ssect:resultshd46}

As mentioned in Section~\ref{sssec:hd46703}, the inclination angle of the binary system in HD\,46703 is most likely smaller than $60\degr$. Hence, we set this angle as the upper limit for the spatio-kinematic modelling. The jet half-opening angle can take on values between $10\degr$ and $80\degr$. The resulting best-fitting model parameters are tabulated in Table~\ref{tab:bestfit} and the geometry is represented in Figure~\ref{fig:geom_hd}. The jet in HD\,46703 has an opening angle of $27\degr$ and a tilt of $5\degr$ with respect to the orbital axis of the binary system. The outflow velocities are rather low, starting at $0.1\,$\kms at the jet edges and reaching up to $177\,$\kms at the jet centre. Just as for EP\,Lyr, the jet half-opening angle is smaller than the inclination angle of $31\degr$. 

Since the stellar jet configuration fits the data best, we use it as input for the radiative transfer modelling. The jet temperatures in our grid range from $4300\,$K up to $6300\,$K in steps of $100\,$K and the logarithmic jet densities from 14 up to 18 in steps of 0.2. The best-fitting jet temperature and density for HD\,46703 are $T=5400^{+200}_{-1300}\,$K and $\log\rho = 16.1^{+0.8}_{-1.5}$ (see Figure~\ref{fig:chisq_hd}). The resulting jet mass-loss rate that we find is $8\times10^{-8}\,$\myr. This is the lowest mass-loss rate for this sample. The one-sigma confidence interval ranges from $3\times10^{-9}\,$\myr\, to $5\times10^{-7}\,$\myr.


\subsection{Results: HP Lyr}\label{ssect:resultshplyr}
\begin{figure}
\centering
\includegraphics[width=.45\textwidth]{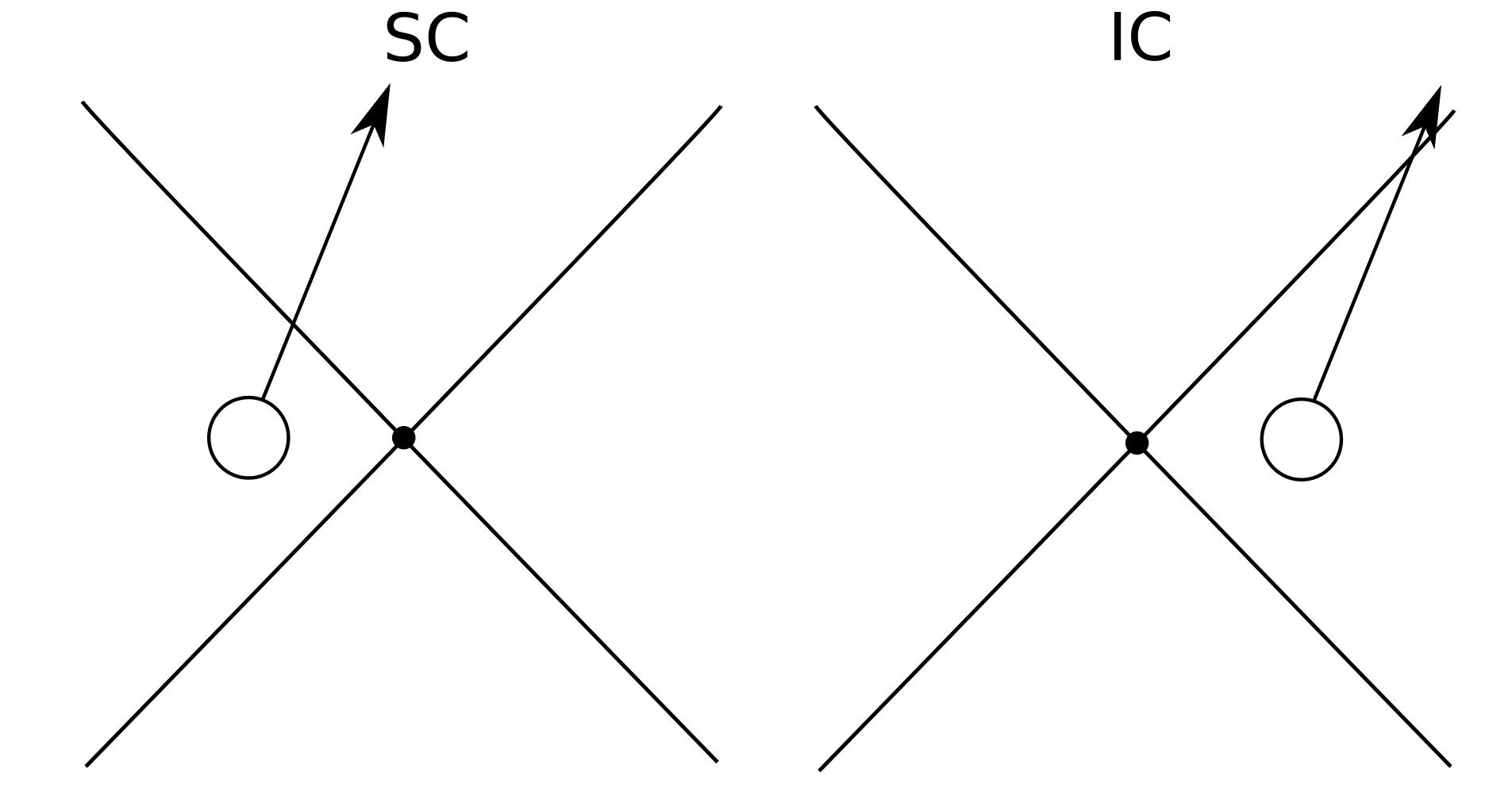}
\caption{Illustration of a binary-jet system where the viewing angle is smaller than the jet half-opening angle. The primary component is represented as a circle. The jet is represented by a double cone, which is centred on the secondary star (small full circle). The viewing direction towards the observed is indicated by the arrow. The left image illustrates the system during superior conjunction (SC) and the right image illustrates the system during inferior conjunction (IC).}\label{fig:viewingangle}
\end{figure}
For the spatio-kinematic modelling of HP\,Lyr, we did not set strict limits on the model parameters. The best-fitting configuration for the jet is an X-wind, with a very wide half-opening angle of $63\degr$ and a jet tilt of almost $10\degr$. The density in the jet peaks at a jet angle of $34\degr$ and is rather uniform throughout the jet, except for the inner jet region ($<\,25\degr$) where the jet has a cavity. The jet velocities range from $11\,$\kms up to $211\,$\kms.

The shape of the absorption feature in HP\,Lyr is well reproduced by the model, as can be seen in Figure~\ref{fig:obsmod_hplyr}. The best-fitting model for HP\,Lyr finds a jet half-opening angle that is larger than the inclination angle of the binary system, which was also the case for 89\,Her. However, contrary to 89\,Her, the observed absorption feature is not present over the whole orbital period, but covers about half of the orbital period. This raises the question as to how it is possible that our line-of-sight towards the primary is always blocked, yet the continuum light in \halpha\,is not always scattered by the jet. The reason for this feature is most-likely the location where the photospheric light of the primary, that travels to the observer, enters the jet. We have illustrated this in Figure~\ref{fig:viewingangle}, where we show a jet-binary system with an inclination angle, or viewing angle, smaller than the jet half-opening angle. During superior conjunction, the photospheric light will enter the jet closer to its base, where the density is high. During inferior conjunction, the companion and primary star have switched positions. Now, the photospheric light will pass through the jet at a greater height. Hence, the entry point will be further from the base, where the density is significantly smaller compared to the entry point at superior conjunction. This means that the amount of scattering of photospheric light by the jet is too low to observe an absorption feature in \halpha. 

The grid of jet temperatures and densities for the radiative transfer modelling is set from $4000\,$K to $6000\,$K in steps of $100\,$K and from 14 to 18 in steps of 0.2, respectively. Here, we find a jet temperature $T=5400^{+200}_{-900}\,$K and a jet density $\log\rho = 16.6^{+0.9}_{0.3}$ (see Figure~\ref{fig:chisq_hplyr}). The corresponding jet mass-loss rate for HP\,Lyr is $1.6\times10^{-6}\,$\myr, with a one-sigma confidence interval from $8\times10^{-7}\,$\myr\, to $1.3\times10^{-5}\,$\myr.


\subsection{Results: TW Cam}\label{ssect:resultstwcam}

As for HP\,Lyr, we did not set strict limits on the model parameters for the spatio-kinematic modelling of TW\,Cam. For this object, we find the best-fitting jet configuration to be an X-wind, which is represented in Figure~\ref{fig:geom_twcam}. The binary system has a low inclination angle of $26\degr$ and the post-AGB star has a large radius of $83\,$R$_\odot$. The jet has a half-opening angle of $49\degr$ with a jet cavity of $25\degr$, and is tilted by $5\degr$ with respect to the orbital axis of the binary system. The jet in TW\,Cam has the highest outflow velocity in this sample with $v_\text{out} = 443\,$\kms.

The corresponding dynamic \halpha\,line for the best-fitting model of TW\,Cam is shown in Figure~\ref{fig:obsmod_twcam}. The model can successfully reproduce the complex double-peaked absorption feature in the line. The model predicts an inclination angle ($26\degr$) that is smaller than the jet half-opening angle, even though the absorption is not observed throughout the whole orbital period. Hence, we assume the same argument for TW\,Cam as for HP\,Lyr, that the scattering by the jet is not substantial enough to create an absorption feature during inferior conjunction.

For the radiative transfer modelling, we set a range from $3600\,$K to $5000\,$K in steps of $100\,$K for the jet temperature and from 17 to 21 in steps of 0.2 for the logarithm of the jet density. We only fit the EW of the \halpha\,and \hbeta\, from our model to the observations, since the uncertainties in the higher Balmer lines become too large. The best-fitting jet temperature and density for TW\,Cam are $T=4400^{+200}_{-400}\,$K and $\log\rho = 18.4^{+0.9}_{-0.2}$ (see Figure~\ref{fig:chisq_twcam}). This translates to a jet mass-loss rate of $1.8\times10^{-4}\,$\myr, which is the highest mass-loss rate for the jets in our sample. The one-sigma confidence interval ranges from $1.1\times10^{-4}\,$\myr\, to $1.4\times10^{-3}\,$\myr. 


\section{Discussion}\label{sec:discussion}


\subsection{The nature of the jet}\label{ssec:nature}


\subsubsection{The jet-launching mechanism}\label{sssec:launch}

For the spatio-kinematic modelling of the jet, we fit three distinct jet configurations representing three different jet-launching mechanisms: the stellar jet, the X-wind, and the disc-wind configurations.

The reduced chi-square values for the fitting are listed in Table~\ref{tab:bestfit} and the dynamic \halpha\, model spectra are shown in Figure~\ref{fig:allmodels}. For 89\,Her, EP\,Lyr, and HD\,46703, the stellar jet configuration gives the lowest chi-square value. For HP\,Lyr and TW\,Cam, the X-wind configuration provides a better fit. This said, the difference in chi-square between the three jet configurations is not always statistically significant. The main reason for this is that the line-of-sight probes the jet at a height of $1-2\,$AU from the jet launch-point, while the size of the launching region is a factor of $\sim 100$ smaller than this height. At these heights, it becomes difficult to distinguish between the three jet configurations. This can be seen in Figure~\ref{fig:geom}. For example, the stellar jet configurations of EP\,Lyr and HD\,46703 have a low density region along the jet axis. This mimics the jet cavity that is included in the X-wind and disc-wind configurations, such that this jet resembles the X-wind of TW\,Cam. {\it Hence, based on our fitting results, we cannot easily differentiate between the different jet-launching mechanisms. }


\subsubsection{Jet angle and velocity}\label{sssec:jetanglevel}

The jets in all five systems have wide half-opening angles between $\sim 30 - 60\,$\degr. These wide jets are commonly observed in other evolved binary systems and can explain the observed shapes of PNe and proto-PNe \citep{soker04, akashi08,akashi13, akashi18}, which are closely related to our post-AGB binary jets. Additionally, the maximum jet velocities, which range from $150$ to $440\,$\kms, are of the order of the escape velocity of a main sequence star, rather than, for example, a white dwarf. This strengthens the argument by \cite{bollen17} that the companion star in these systems is most likely always a main sequence star.


\subsubsection{Jet-binary plane misalignment}\label{sssec:jettilt}

\begin{table*} 
    \begin{center}
    \caption{Orbital parameters of 89\,Her, EP Lyr, HD 46703, HP Lyr, and TW Cam from \citet{oomen18} and this work (\textit{last four columns}). The last four columns are the orbital separation $a$, the mass of the secondary $M_2$, the mass ratio $q=M_1/M_2$, and the Roche-lobe filling factor $f=R_1/R_\text{RL,1}$, based on the orbital parameters and the results of the spatio-kinematic modelling. We assume that $M_1=0.6\,$M$_\odot$.}
    \label{tab:stelorbpar}
    \scalebox{.8}{\begin{tabular}{l c c c c c c c c c c c c c}
        \hline
        Object     & Period         & Eccentricity  & $T_0$             & $\omega$     & $K_1$        & $\gamma$       & $a_1\sin i$     & $f(m)$            & $a$  & $M_2$     & $q$ & $f$ \\
                   & days           &               & days              & Degrees      & km$\,s^{-1}$ & km$\,s^{-1}$   & AU              & M$_\odot$         &  AU  & M$_\odot$ &             &  \\
        \hline
        89\,Her    & $289.1\pm0.2$  & $0.29\pm0.07$ & $2447832\pm12$    &  $68.4\pm15$ & $4.2\pm0.3$  & $-27.0\pm0.2$  & $0.106\pm0.007$ & $0.0019\pm0.0004$ & 1.07 & 1.36      & 0.44        & 0.69 \\
        EP\,Lyr    & $1151\pm14$    & $0.39\pm0.09$ & $2455029.3\pm8.7$ & $61.1\pm7.6$ & $13.4\pm1.3$ & $15.9\pm1.0$   & $1.30\pm0.12$   & $0.22\,0.06$      & 2.5  & 0.95      & 0.63        & 0.4 \\
        HD\,46703  & $597.4\pm0.2$  & $0.30\pm0.02$ & $2443519.6\pm7.6$ & $61.9\pm4.2$ & $16.0\pm0.3$ & $-93.3\pm0.2$  & $0.839\pm0.015$ & $0.220\pm0.012$   & 2.02 & 2.45      & 0.24        & 0.67 \\
        HP\,Lyr    & $1818\pm80$    & $0.20\pm0.04$ & $2456175\pm61$    & $14\pm13$    & $7.8\pm0.2$  & $-115.6\pm0.2$ & $1.27\pm0.06$   & $0.083\pm0.007$   & 4.66 & 3.5       & 0.17        & 0.31 \\
        TW\,Cam    & $662.2\pm5.3$  & $0.25\pm0.04$ & $2455111\pm18$    & $144\pm10$   & $14.1\pm0.6$ & $-49.8\pm0.5$  & $0.83\pm0.04$   & $0.174\pm0.022$   & 2.29 & 3.05      & 0.20        & 0.67 \\
        \hline
    \end{tabular}}
    \end{center}
\end{table*}

The set-up of the jet in our model is such that the jet axis can be inclined with respect to the rotation axis of the binary system. A positive jet tilt corresponds to a jet for which the cone through which the line-of-sight passes through is tilted away from the direction of travel of the companion. An interesting outcome of this study is the range in jet inclinations or tilts that we find in our sample, which varies from $-5\degr$ up to $10\degr$. The jet tilt has a strong effect on the \halpha\, line absorption feature. A positive jet tilt will cause the jet absorption to be observed later in phase. Hence, the peak of the absorption feature will not occur during superior conjunction, but during a later orbital phase.

A tilted jet is a common phenomenon in jet-launching systems. Many jets show a long-term precessing motion. Examples are the precessing jets in protostellar system V380 Ori NE \citep{choi17} and in the "water fountain" evolved systems IRAS16342$-$3814 \citep{sahai17} and IRAS 18286$-$0959 \citep{yung11}. The jet in IRAS 18286$-$0959 reaches a precession half-opening angle of about $14\degr$, showing that large jet tilts are possible. 

The three-dimensional magneto-hydrodynamic simulations performed by \cite{Sheikhnezami15} showed that a jet tilt might be connected to binarity. They simulated a jet launched from a star surrounded by an accretion disc, in a single star system and in a binary system. For the simulations of a jet launched by the star in a binary system, they found jet bending and the onset of precession, with a precession angle of $4\degr$. In a later study by \cite{Sheikhnezami18}, similar simulations were performed for different binary mass ratios, binary separations, and inclination angles between the accretion disc and the orbital plane. They assumed mass ratios $q=M_\text{d}/M_\text{a}=$ 0.5, 1, and 2, with $M_\text{d}$ and $M_\text{a}$ the mass of the donor star and the accretor, respectively. The orbital separations in their simulations are $a=150\,R_\text{in}$ and $200\,R_\text{in}$, with $R_\text{in}$ the inner disc radius of the accretion disc, and accretion disc inclinations of $0\degr$, $10\degr$, and $30\degr$. In their simulations, the offset of the jet rotation axis relative to its initial position reached $8\degr$ for the model with $q=1$, $a=150\,R_\text{in}$, and a jet inclination of $\delta=10\degr$. The mass ratios of the binaries in our sample range from 0.17 up to 0.63, with orbital separations ranging from $62\,R_\text{in}$ to $187\,R_\text{in}$ (assuming an inner disc radius that is three times the stellar radius of the companion star). These parameters are listed in Table~\ref{tab:stelorbpar}. These orbital and stellar parameters are similar to the model setup of \cite{Sheikhnezami18}. This is especially the case for EP\,Lyr, with $q=0.63$ and $a = 187\,R_\text{in}$. Hence, these simulations and the aforementioned observations of precessing jets lead us to assume that the jets in our sample are similar in nature. 

If the jets in our sample have a precessing motion, the jet absorption feature in \halpha\,will shift as a function of orbital phase over time. According to \cite{bate00}, the precession period of the accretion disc, and thus the jet, will be about 20 times the orbital period of the binary. The precession periods for the objects in our sample would range from 16 years to 100 years.


\subsection{Source feeding the accretion}

\begin{table*}
\begin{center}
\caption{Derived accretion and mass-transfer rates in the binary systems. The tabulated parameters are: the mass-loss rate of the jet $\dot{M}_\text{jet}$, the mass-accretion rate in the circum-companion accretion disc $\dot{M}_\text{accr}$, the mass-transfer rate from the circumbinary disc $\dot{M}_\text{tr,CBD}$, and the mass-transfer rate from the post-AGB star $\dot{M}_\text{tr,pAGB}$. For each parameter, we show the average value and its upper and lower bound.}
\label{tab:mrates}
\scalebox{.83}{\begin{tabular}{l ccc ccc ccc ccc}\\
\hline
Object & \multicolumn{3}{c}{$\dot{M}_\text{jet}$} &  \multicolumn{3}{c}{$\dot{M}_\text{accr}$} &  \multicolumn{3}{c}{$\dot{M}_\text{tr,CBD}$} & \multicolumn{3}{c}{$\dot{M}_\text{tr,pAGB}$}  \\
       & \multicolumn{3}{c}{\myr} &  \multicolumn{3}{c}{\myr} &  \multicolumn{3}{c}{\myr} & \multicolumn{3}{c}{\myr}  \\
       \cmidrule(lr){2-4} \cmidrule(lr){5-7}  \cmidrule(lr){8-10} \cmidrule(lr){11-13}
        &  Lower & Average & Upper   &  Lower & Average & Upper    &  Lower & average & Upper & Lower & Average & Upper\\       
\hline 
89\,Her   & $1.4\times10^{-6}$ & $2\times10^{-6}$   & $9\times10^{-6}$   & $7\times10^{-6}$   & $1\times10^{-5}$ & $5\times10^{-5}$ & $3\times10^{-8}$   & $5\times10^{-7}$   & $3\times10^{-6}$   & $1.6\times10^{-9}$  &  $1.6\times10^{-8}$ &  $8\times10^{-8}$\\
EP\,Lyr   & $9\times10^{-9}$   & $9\times10^{-8}$   & $2\times10^{-6}$   & $5\times10^{-8}$   & $5\times10^{-7}$ & $1\times10^{-5}$ & $1.7\times10^{-8}$ & $3\times10^{-7}$   & $1.5\times10^{-6}$ & $3\times10^{-9}$  & $3\times10^{-8}$  & $1.6\times10^{-7}$  \\
HD\,46703 & $3\times10^{-9}$   & $8\times10^{-8}$   & $5\times10^{-7}$   & $1.5\times10^{-8}$ & $4\times10^{-7}$ & $3\times10^{-6}$ & $1.6\times10^{-8}$ & $3\times10^{-7}$   & $1.3\times10^{-6}$ & $4\times10^{-9}$  &  $4\times10^{-8}$ &  $2\times10^{-7}$  \\
HP\,Lyr   & $8\times10^{-7}$   & $1.6\times10^{-6}$ & $1.3\times10^{-5}$ & $4\times10^{-6}$   & $8\times10^{-6}$ & $7\times10^{-5}$ & $6\times10^{-9}$   & $1.0\times10^{-7}$ & $5\times10^{-7}$   & $4\times10^{-9}$   & $4\times10^{-8}$ &  $2\times10^{-7}$  \\
TW\,Cam   & $1.1\times10^{-4}$ & $1.8\times10^{-4}$ & $1.4\times10^{-3}$ & $6\times10^{-4}$   & $9\times10^{-4}$ & $7\times10^{-3}$ & $1.0\times10^{-8}$ & $1.7\times10^{-7}$ & $8\times10^{-7}$   & $6\times10^{-10}$  & $6\times10^{-9}$  &  $3\times10^{-8}$  \\
\hline
\end{tabular}}
\end{center}
\end{table*}

An important question to understand the formation and evolution of post-AGB binaries is the nature of the source feeding the circum-companion accretion disc. The plausible sources are the post-AGB star, the circumbinary disc or both. In \cite{bollen20}, we calculated the accretion rates in the circum-companion accretion disc based on the mass-loss rates in the jet and compared these values with estimated mass-loss rates from the circumbinary disc and the post-AGB star. We showed that the circumbinary disc is more likely to be the source feeding the accretion. However, mass transfer from both the circumbinary disc and the post-AGB star cannot be excluded. Here, we will perform a similar analysis, in order to determine the source feeding the circum-companion accretion disc. 


\subsubsection{Mass-transfer rates}

The estimation of the mass-accretion rates onto the companion is dependent on the mass-ejection rates in the jet. Here, we assume an ejection efficiency ($\dot{M}_\text{jet, tot}/\dot{M}_\text{accr}$)\footnote{$\dot{M}_\text{jet,tot}$ refers to the mass-loss rate from both jet lobes. The mass-loss rates calculated from the model are those for one lobe.} of 0.4, but we note that this figure is very much an estimate and that values in the range $0.04$ to $1$ are possible \citep{shu94, ferreira06, blackman14}. The resulting accretion rates for the jets in our sample are listed in Table~\ref{tab:mrates}. 

First, we compute the possible mass transfer from the stellar wind of the post-AGB star to the accretion disc around the companion. When a star is in the post-AGB phase, most of the envelope will already have been removed, and the star experiences a decreased mass loss with typical mass-loss rates of the order of $10^{-9}\,-10^{-7}$\myr \citep{blocker95, bertolami19}. However, there are still large uncertainties associated with determining the mass-loss rates of stellar winds in post-AGB stars \citep{soker02b, vanwinckel03, cranmer11}. Here, we will use the prescription by \cite{schroder05}, which is based on Reimers law \citep{reimers75}:
\begingroup\small
\begin{equation}
    \dot{M}_\text{pAGB} = 8\times 10^{-14} M_\odot/\text{yr} \left(\frac{M_1}{M_\odot} \right)^{-1} \left(\frac{L_1}{L_\odot} \right) \left( \frac{R_1}{R_\odot}\right) \left( \frac{T_\text{eff}}{4000\,\text{K}}\right)^{3.5} \left( 1 + \frac{g_\odot}{4300\,g_1}\right), \label{eq:schroder}
\end{equation}
\endgroup
with $M_1$, $L_1$, $R_1$, $T_\text{eff}$, and $g_1$ the mass, luminosity, radius, effective temperature, and surface gravity of the primary. The actual mass-transfer rate from the post-AGB star to the companion is highly dependent on the mass-transfer method, i.e., Roche-lobe overflow (RLOF), wind-RLOF \citep{mohamed07}, or Bondi-Hoyle-Lyttleton (BHL) accretion \citep{hoyle39, bondi44, edgar04}. For RLOF to occur, the post-AGB star has to fill its Roche lobe. However, the Roche-lobe filling factor of these objects lies between 0.3 and 0.7. Hence, we can rule out this method. Instead, we will assume that mass is being accreted by the companion via wind-RLOF or BHL accretion. The efficiency of wind-RLOF and BHL accretion highly depend on the stellar and orbital parameters of the binary system \citep{edgar04, abate13}. Most simulations show that the mass-transfer efficiency ($\dot{M}_\text{accr}/\dot{M}_\text{pAGB}$) ranges from one percent up to fifty percent \citep{mohamed07, abate13, saladino19, chen20}. Hence, we assume this range as well for the possible mass-transfer efficiency for the post-AGB wind. We list the resulting mass-transfer rates from the post-AGB star ($\dot{M}_\text{tr,pAGB}$) for a mass-transfer efficiency of $1\%$ (lower), $10\%$ (average), and $50\%$ (upper) in Table~\ref{tab:mrates}.

Next, we estimate the re-accretion rate from the circumbinary disc onto the central binary system. In order to calculate the mass that is being accreted by the binary system, we follow the prescription by \cite{rafikov16}. We assume the standard $\alpha$-disc model by
\cite{shakura73}, such that the viscosity can be described as
\begin{equation}
	\nu = \frac{\alpha \text{c}_\text{s}^2}{\Omega},\label{eq:viscosity}
\end{equation}
where $\alpha$ is the viscosity parameter, c$_\text{s}= \sqrt{k_B T/\mu}$ is the local speed of sound, and $\Omega$ is the local angular frequency at radius $r$. The local sound speed is a function of the midplane temperature in the disc ($T$). $\mu$ is the mean molecular weight, set to $2\,m_p$. Since the central post-AGB stars in these systems have high luminosities, we assume that the temperature in the disc is governed by the stellar irradiation:
\begin{equation}
	T = \left(\frac{\zeta L_1}{4\pi\sigma}\right)^{1/4} \, r^{-1/2}, \label{eq:discT}
\end{equation}
where $\zeta$ is a constant factor that accounts for the starlight that is intercepted by the disc surface at a grazing incidence angle ($\approx0.1$), $L_1$ is the luminosity of the post-AGB star and $\sigma$ is the Stefan-Boltzmann constant. The viscous time of the disc is defined as $t_0=r^2/\nu$. By implementing the viscosity from Equation~\ref{eq:viscosity} and using the disc temperature defined in Equation~\ref{eq:discT}, the viscous time becomes
\begin{equation}
	t_0 = \frac{4}{3}\frac{\mu}{k_\text{B}}\frac{a}{\alpha} \left[ \frac{4\pi\sigma(GM_b)^2}{\zeta L_1} \right]^{1/4} \left( \frac{\eta}{I_L}, \right)^2\label{eq:t0}
	,\end{equation}
where $m_p$ is the mass of a proton, $M_b$ is the total mass of the binary, $\eta$ is the ratio of angular momentum of the disc compared to that of the central binary, and $I_L$ ($=1.8$) a dimensionless factor that characterises the distribution of the angular momentum in the disc \citep{rafikov16b}. For the viscosity parameter $\alpha$, we use a value of $10^{-2}$, which is a typical value found for the circumbinary discs in post-AGB binaries by \citet{oomen20}. The parameter $\eta$ is set to 2, which resembles a circumbinary disc for which the surface density scales with $r^{-3/2}$. The viscous time of the disc, defined in Equation~\ref{eq:t0}, gives a good estimate for the disc lifetime. For the objects in our sample, this results in disc lifetimes on the order of $10^3 - 10^4$ years. These values are consistent with typical lifetimes of post-AGB stars, which last for $10^3-10^5$ years \citep{miller16}.

The mass loss from the disc will decrease over time and is dependent on the disc mass, the binary parameters, and the stellar parameters of the system: 
\begin{equation}
	\dot{M}_\text{tr, CBD}(t) = \frac{1}{2}\frac{M_{0,\text{CBD}}}{t_0} \left( 1 + \frac{t}{2t_0} \right)^{-3/2},\label{eq:masslossdisc}
\end{equation}
where $M_{0,\text{CBD}}$ is the initial disc mass and $t$ is time. We assume $t=0$ and include a factor of $1/2$, since we assume that the re-accreted mass is equally divided between the two stars in the binary system. Observations show that circumbinary discs around post-AGB stars have typical disc masses between $6\times10^{-4}\,\text{M}_\odot$ and $5\times 10^{-2}\,\text{M}_\odot$ \citep{gielen07,bujarrabal13a,bujarrabal18,hillen17, kluska18}. We will adopt these values to calculate the lower and upper bounds for the re-accretion rate. The resulting estimated re-accretion rate from the circumbinary disc onto the companion ($\dot{M}_\text{tr,CBD}$) and its limits are listed in Table~\ref{tab:mrates}. For the average re-accretion rate, we assume a disc mass of $10^{-2}\,\text{M}_\odot$.

\begin{figure}
\centering
\includegraphics[width=.5\textwidth]{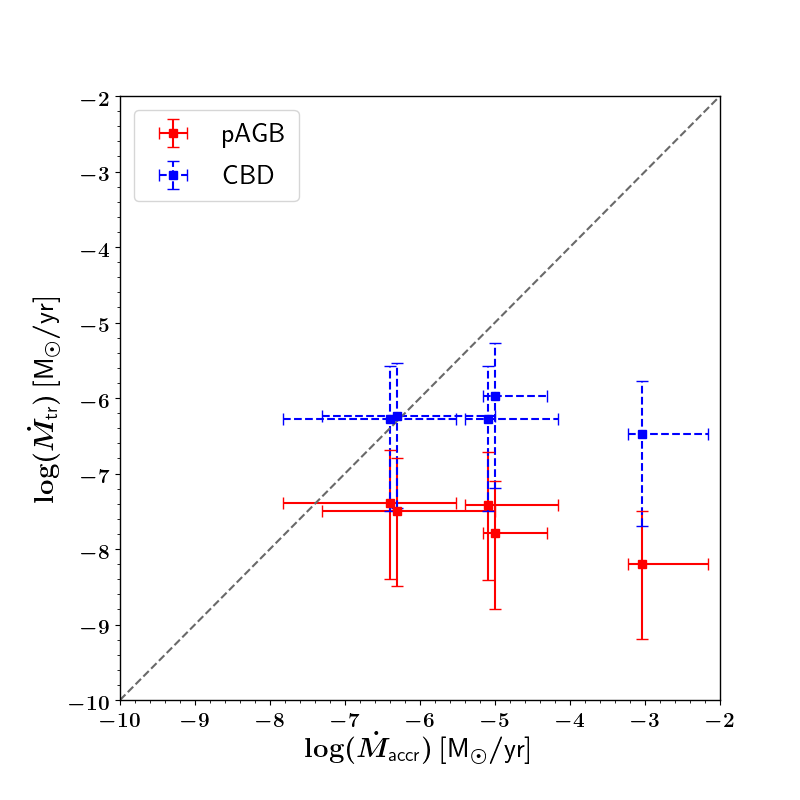}
\caption{The estimated mass-transfer rates to the companion from both the post-AGB star (\textit{full red}) and the circumbinary disc (\textit{dashed blue}) vs. the mass-accretion rates onto the companion derived from the model for the five objects. The values are given in logarithmic scale and the grey dashed line represents the identity line.}\label{fig:mratesvsmrates}
\end{figure}


\subsubsection{Comparison of mass-accretion and mass-transfer rates}

We compare the mass-transfer rates from the post-AGB star ($\dot{M}_\text{tr,pAGB}$) and the circumbinary disc ($\dot{M}_\text{tr,CBD}$) onto the companion with the mass-accretion rates of the circum-companion disc in Figure~\ref{fig:mratesvsmrates}. For EP\,Lyr and HD\,46703, the mass-transfer rate from the circumbinary disc is in good agreement with the mass-accretion rate of the companion. Additionally, the accretion models of EP\,Lyr in \citet{oomen20} reach similar re-accretion rates of the circumbinary disc onto the central binary of the order of $10^{-7}\,$\myr. This is a strong indication for this object that re-accretion is the source feeding the circum-companion accretion disc. The final re-accretion rates found by \citet{oomen20} for HD\,46703 are one order of magnitude lower ($0.5-5 \times 10^{-8}\,$\myr), but still fall within the uncertainties.

Figure~\ref{fig:mratesvsmrates} shows that for each object the estimated mass-transfer rates from the circumbinary disc is higher than the estimated mass-transfer rate from the post-AGB star. Since the mass must be provided by at least one of these sources, it is thus more likely that the circumbinary disc provides the bulk of the mass to the companion.

For TW\,Cam the mass-accretion rate is three orders of magnitude larger than the expected mass-transfer rate from the circumbinary disc and five orders of magnitude larger than the mass-transfer rate from the post-AGB star. This high mass-accretion rate of $9\times10^{-4}\,$\myr is improbable, since it would imply that either the post-AGB star has a mass-loss rate $>10^{-3}\,$\myr or that the lifetime of the circumbinary disc would be $10$ years (assuming a disc mass of $10^{-2}\,$\myr).

The high mass-accretion rate in TW\,Cam is likely due to the limits of our jet model. In our model, we assume a constant effective temperature for the post-AGB star and a uniform jet temperature.  However, TW\,Cam shows stellar pulsations and shocks, which will cause a significant change in effective temperature. Additionally, since we observe absorption caused by the jet, we can infer that the jet must be colder than the post-AGB star, which is relatively cool ($4800\,$K). The jet density has to be higher at a cooler temperature, in order to observe the absorption feature in the Balmer lines. Hence, the combination of the uniform jet temperature and the constant effective temperature might cause the model to choose a low jet temperature and high densities, which then results in high mass-loss rates. In future work, we will improve on the jet model by incorporating more sophisticated prescriptions to better model the shocks and the jet temperature structure, such that we can better estimate the mass-loss and mass-accretion in such stars.


\section{Summary and conclusions}\label{sec:conclusions}

We have modelled a diverse sample of post-AGB binaries with jets, using the spatio-kinematic and radiative transfer models described in \citet{bollen19} and \citet{bollen20}.

Our sample of five post-AGB binaries comprises a broad range of orbital parameters and different Balmer line variations that are caused by the jet absorption. Our model can successfully reproduce the dynamic \halpha\,spectra for all sources, which demonstrates that it has a sufficient number of physical ingredients to account for the most important features in the Balmer line variability of this diverse group of objects. 

From the spatio-kinematic modelling, we obtained the geometric, kinematic, and relative density structure of the jet. We found jets with wide half-opening angles (between 30 and 60\degr) and jet outflow velocities in the range $150 - 440\,$\kms. Additionally, our results show that a tilted jet is very common for these systems (the range is $5 - 10$\degr). This tilt can be explained by precession. Jet precession is commonly observed in different classes of jet-launching objects \citep[e.g.,][]{choi17}. Moreover, simulations of jets launched by a star in a binary system show that the presence of a companion can cause precessing motion \citep{Sheikhnezami15, Sheikhnezami18}. Jet configurations in our model are based on three jet-launching models: a stellar jet, an X-wind and a disc wind. The best fits assuming each configuration in turn cannot discriminate between the three models. This is mainly caused by the relatively small dimensions of the launching region in all three models, compared to the whole system. Hence, it is not possible to distinguish which process is more likely to be the launching mechanism in these jets. Other observational techniques, such as interferometric imaging with high angular resolution, might be necessary in order to get a better understanding of the jet-launching mechanisms at play in these systems.

From the radiative transfer model, we were able to obtain jet temperatures and densities. By combining the spatio-kinematic structure with the radiative transfer model, we found the jet mass-loss rates for these objects and estimated the mass-accretion rates in the circum-companion accretion disc. We compared these accretion rates with expected mass-transfer rates from the post-AGB star or from the circumbinary disc. We found that the accretion rates are too high to be explained solely by mass-transfer from the post-AGB star to the companion mass-transfer. The accretion rates are in better accordance with a mass-transfer from the circumbinary disc to the companion. We conclude that the circumbinary disc is most likely the main source of the accreted material, however, we cannot exclude the possible mass-transfer from the post-AGB star.

In future work, we will improve upon the jet model, by implementing a more sophisticated temperature structure and radiative transfer calculations. We will also apply improved spatio-kinematic and radiative transfer model to all 15 jet-creating post-AGB binary systems in our target sample. This will provide us with jet parameters, mass-loss rates, and mass-accretion rates. We ultimately aim to correlate jet and accretion properties with the binary properties, which will give us more insights on the binary interaction history of post-AGB binary systems. 


\section*{Acknowledgements}
This work was performed on the OzSTAR national facility at Swinburne University of Technology. OzSTAR is funded by Swinburne University of Technology and the National Collaborative Research Infrastructure Strategy (NCRIS). DK  acknowledges  the  support  of  the  Australian  Research Council  (ARC)  Discovery  Early  Career  Research  Award (DECRA) grant (95213534). DB and HVW acknowledge support from the Research Council of the KU Leuven under grant number C14/17/082.  The observations presented in this study are obtained with the HERMES spectrograph on the Mercator Telescope, which is supported by the Research Foundation - Flanders (FWO), Belgium, the Research Council of KU Leuven, Belgium, the Fonds National de la Recherche Scientifique (F.R.S.-FNRS), Belgium, the Royal Observatory of Belgium, the Observatoire de Gen\`eve, Switzerland and the Th\"uringer Landessternwarte Tautenburg, Germany.

\section*{Data availability}
The data underlying this article will be shared on reasonable request to the corresponding author.


\bibliographystyle{mnras}
\bibliography{allreferences} 


\appendix


\section{Results from the spatio-kinematic modelling}

\begin{table*} 
    \begin{center}
    \caption{Best-fitting jet parameters for the spatio-kinematic model of all three configurations for the five objects. We also include the reduced chi-square values for the three jet configurations. The tabulated parameters are similar as those in Table~\ref{tab:bestfit}. $c_v$ is the scaling parameter for the outer jet velocity of the disc-wind configuration.}
    \label{tab:allbestfit}
    \scalebox{.9}{\begin{tabular}{l ccc ccc ccc}
        \hline
                  & \multicolumn{3}{c}{89\,Her} & \multicolumn{3}{c}{EP Lyr} & \multicolumn{3}{c}{EP Lyr - puls} \\
       \cmidrule(lr){2-4} \cmidrule(lr){5-7}  \cmidrule(lr){8-10} 
        Parameter                           &     stellar       &    X-wind       &     disc-wind  &     stellar       &  X-wind       &  disc-wind     &     stellar      &    X-wind    &     disc-wind \\
        \hline
         $i$ ($\degr$)                      &   $8.2\pm0.2$     &  $8.1\pm0.9$    &  $8.1\pm0.1$   &  $58.2\pm0.3$     &  $75.1\pm0.9$ &  $62\pm0.5$    &  $58.0\pm0.3$    &  $58\pm2$       &  $60.3\pm0.4$  \\
         $\theta_\text{out}$ ($\degr$)      &   $31.1\pm0.1$    &  $38.6\pm0.9$   &  $77.7\pm0.8$  &  $36.0\pm0.3$     &  $69\pm2$     &  $50.9\pm0.9$  &  $35.4\pm0.4$    &  $56\pm2$       &   $80.3\pm0.2$ \\
         $\theta_\text{in}$ ($\degr$)       &                   &  $34.0\pm0.4$   &  $32.6\pm0.5$  &                   &  $40.0\pm1.2$ &  $42.2\pm0.6$  &                  &  $45\pm3$       &  $40.8\pm0.6$  \\
         $\theta_\text{cav}$ ($\degr$)      &                   &  $25\pm0.3$     &  $25\pm1$      &                   &  $25.0\pm0.9$ &  $25.1\pm0.2$  &                  &  $29\pm2$       &  $25.7\pm0.6$  \\
         $\phi_\text{tilt}$ ($\degr$)       &   $-1.4\pm0.3$    &  $0.6\pm0.3$    &  $14.9\pm0.9$  &  $9.3\pm0.2$      &  $4.8\pm1.3$  &  $10.3\pm0.3$  &  $9.0\pm0.3$     &  $-6.1\pm0.9$   &  $11.9\pm0.4$  \\
         $v_\text{in}$ (km\,s$^{-1}$)       &   $155\pm1$       &  $177\pm4$      &  $160\pm20$    &  $280\pm5$        &  $590\pm40$   &  $243\pm3$     &  $300\pm9$       &  $780\pm50$     &  $237\pm3$  \\
         $v_\text{out}$ (km\,s$^{-1}$)      &   $0.15\pm0.10$   &  $0.3\pm0.3$    &  $75\pm9$      &  $32.3\pm1.3$     &  $71\pm3$     &  $128\pm2$     &  $32\pm2$        &  $53\pm3$       & $73\pm6$  \\
         $c_v$                              &                   &                 &  $0.01\pm0.01$ &                   &               &  $0.19\pm0.02$ &                  &                 &  $0.13\pm0.02$  \\
         $p_\text{v}$                       &   $-0.41\pm0.06$  &  $-0.08\pm0.15$ &  $7\pm2$       &  $-2.9\pm0.2$     &  $8.4\pm0.3$  &  $5\pm2$       &  $-2.5\pm0.2$    &  $0.68\pm0.08$  &  $1\pm2$  \\
         $p_{\rho\text{,in}}$               &                   &  $14.4\pm0.5$   &  $14.1\pm0.7$  &                   &  $10.7\pm0.9$ &  $11.6\pm0.6$  &                  &  $14.7\pm0.7$   &  $13.9\pm0.7$  \\
         $p_{\rho\text{,out}}$              &   $1.88\pm0.07$   &  $1.8\pm1.1$    &  $-10.9\pm0.9$ &  $14.8\pm0.4$     & $14.7\pm0.9$  &  $10\pm2$      &  $14.9\pm0.4$    &  $14.7\pm0.1$   &  $-13.5\pm0.3$  \\
         $c_\tau$                           &   $2.67\pm0.01$   &  $2.7\pm0.2$    &  $3.3\pm0.2$   &  $2.82\pm0.03$    &  $0.7\pm0.2$  &  $2.65\pm0.06$ &  $2.89\pm0.02$   &  $1.7\pm0.3$    &  $3.05\pm0.07$  \\
         $R_\text{1}$ (AU)                  &   $0.231\pm0.001$ &  $0.21\pm0.2$   &  $0.22\pm0.02$ &  $0.341\pm0.005$  &  $0.32\pm$    &  $0.34\pm0.01$ &  $0.32\pm0.01$   &  $0.08\pm0.05$  &  $0.34\pm0.02$  \\
         \hline
         $\chi^2_{\nu}$                     &   $0.3$           &   $0.7$         &     $0.7$      &  $1.7$            &  $3.4$  &   $2.1$ & $1.7$            &   $8.2$   &   $2.2$    \\
        \hline
        \hline
                        & \multicolumn{3}{c}{HD 46703} & \multicolumn{3}{c}{HP Lyr} & \multicolumn{3}{c}{TW Cam} \\
        \cmidrule(lr){2-4} \cmidrule(lr){5-7}  \cmidrule(lr){8-10}
                                           &     stellar             & X-wind           &     disc-wind  &     stellar       &    X-wind    &     disc-wind       &     stellar    &    X-wind    &     disc-wind  \\
        \hline
        $i$ ($\degr$)                      &         $31.2\pm0.3$    &  $59.9\pm0.1$   &  $59.9\pm0.3$   &  $24.7\pm0.2$     &  $18.6\pm0.7$     &  $24.6\pm0.3$   &  $25.7\pm0.1$    &  $25.7\pm0.2$     &  $25.8\pm0.1$ \\
        $\theta_\text{out}$ ($\degr$)      &         $27.1\pm0.3$    &  $57.6\pm0.2$   &  $53.1\pm0.5$   &  $34.5\pm0.2$     &  $62.8\pm0.8$     &  $78.6\pm0.7$   &  $31.2\pm0.1$    &  $48.8\pm0.8$     &  $73\pm1$ \\
        $\theta_\text{in}$ ($\degr$)       &                         &  $54.6\pm0.2$   &  $53.0\pm0.4$   &                   &  $34.2\pm0.5$     &  $34.1\pm0.4$   &                   &  $33.1\pm0.3$    &  $38.1\pm0.2$ \\
        $\theta_\text{cav}$ ($\degr$)      &                         &  $20\pm2$       &  $21.2\pm0.9$   &                   &  $25.0\pm0.3$     &  $25.3\pm0.3$   &                   &  $25.0\pm0.1$    &  $25.3\pm0.2$ \\
        $\phi_\text{tilt}$ ($\degr$)       &         $-5.00\pm0.01$  &  $-0.1\pm0.2$   &  $-0.1\pm0.2$   &  $3.29\pm0.07$     &  $9.6\pm0.3$      &  $13.4\pm0.3$   &  $2.92\pm0.04$    &  $4.9\pm0.1$     &  $5.3\pm0.2$ \\
        $v_\text{in}$ (km\,s$^{-1}$)       &         $177\pm6$       &  $290\pm30$     &  $188\pm6$      &  $176\pm2$        &  $211\pm3$        &  $188\pm3$      &   $700\pm20$      &  $443\pm4$       &  $530\pm6$ \\
        $v_\text{out}$ (km\,s$^{-1}$)      &         $0.11\pm0.02$   &  $0.14\pm0.03$  &  $63\pm2$       &  $14.2\pm0.5$     &  $11\pm2$         &  $129\pm9$      &  $30\pm2$         &  $40\pm3$        &  $85\pm5$ \\
        $c_v$                              &                         &                 &  $0.05\pm0.01$  &                   &                   &  $0.43\pm0.03$  &                   &                  &  $0.002\pm0.001$ \\
        $p_\text{v}$                       &         $0.5\pm0.1$     &  $1.2\pm0.2$    &  $-1.2\pm1.2$   &  $-2.00\pm0.05$   &  $8.5\pm0.2$      &  $7.2\pm0.7$    &  $-0.55\pm0.09$   &  $9.9\pm0.1$     &  $4.3\pm0.2$ \\
        $p_{\rho\text{,in}}$               &                         &  $14.7\pm0.4$   &  $14.7\pm0.4$   &                   &  $14.9\pm0.2$     &  $8.3\pm0.4$    &                   &  $8.1\pm0.3$     &  $12.8\pm0.4$ \\
        $p_{\rho\text{,out}}$              &         $4.3\pm0.2$     &  $-14.4\pm0.6$  &  $0\pm1$        &  $2.5\pm0.1$      &  $-12.1\pm0.2$    &  $-6.9\pm0.1$   &  $7.1\pm0.1$      &  $-14.5\pm0.3$   &  $8.1\pm0.4$ \\
        $c_\tau$                           &         $2.05\pm0.02$   &  $1.36\pm0.02$  &  $1.8\pm0.02$   &  $2.75\pm0.01$    &  $3.08\pm0.03$    &  $2.71\pm0.02$  &  $3.22\pm0.02$    &  $2.97\pm0.002$  &  $3.45\pm0.03$ \\
        $R_\text{1}$ (AU)                  &         $0.36\pm0.01$   &  $0.32\pm0.01$  &  $0.32\pm0.01$  &  $0.762\pm0.004$  &  $0.347\pm0.005$  &  $0.74\pm0.01$  &  $0.386\pm0.004$  &  $0.386\pm0.003$ &  $0.381\pm0.005$ \\
          \hline
        $\chi^2_{\nu}$                     &         $1.8$           &   $3.2$ &  $2.0$  &  $3.3$  &  $1.7$            &   $2.4$ &   $1.6$ &  $1.3$           &   $1.7$ \\
           \hline
    \end{tabular}}
    \end{center}
\end{table*}


\section{Results of the equivalent width fitting}

\begin{figure*}
  \centering
    \begin{subfigure}[b]{.4\linewidth}
        \centering\large 
          \includegraphics[width=1\linewidth]{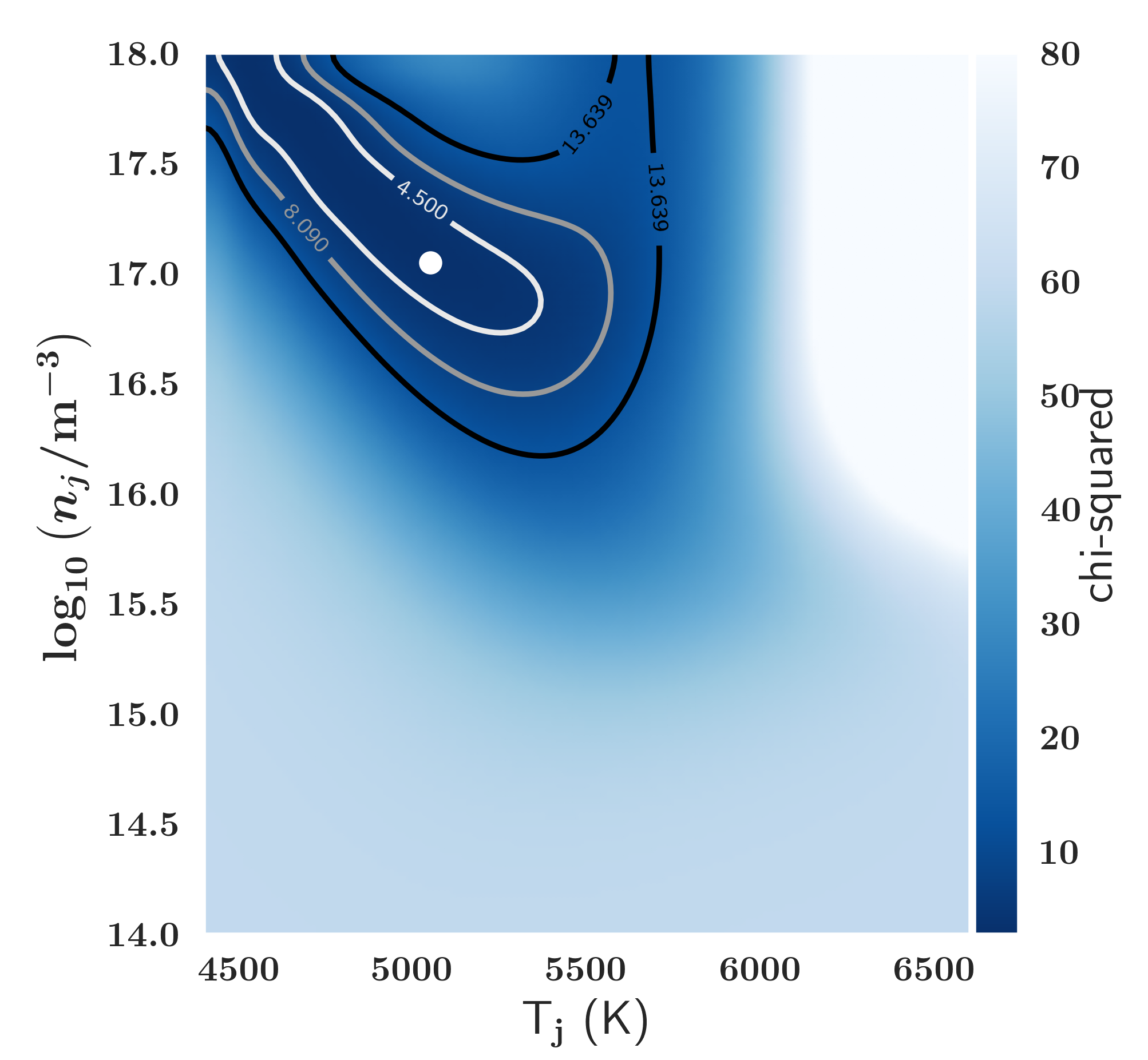}
        \caption{89\,Her}\label{fig:chisq_89}
    \end{subfigure}%
    \begin{subfigure}[b]{.4\linewidth}
        \centering\large 
          \includegraphics[width=1\linewidth]{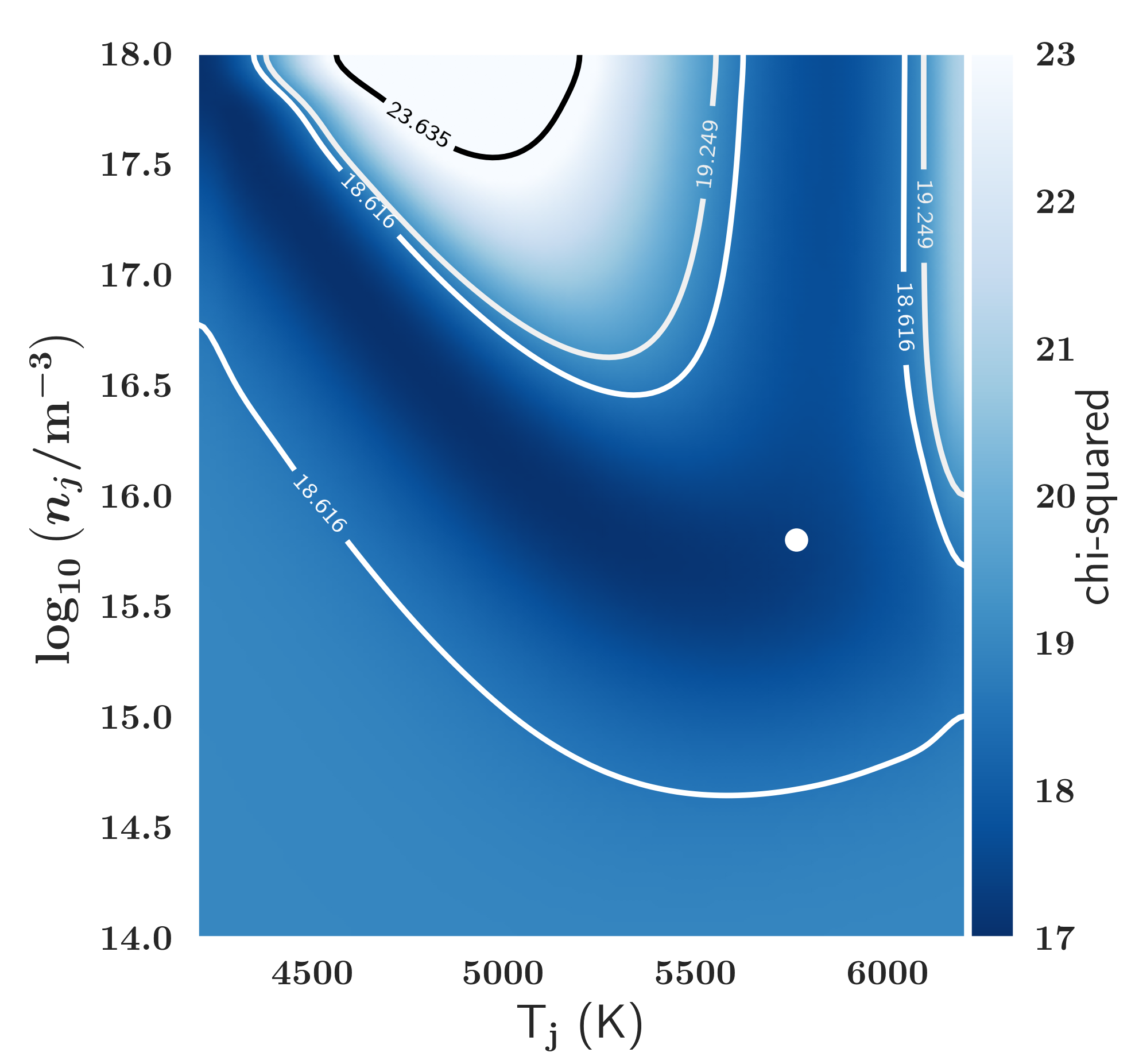}
        \caption{EP Lyr}\label{fig:chisq_eplyr}
    \end{subfigure}%

    \begin{subfigure}[b]{.4\linewidth}
        \centering\large 
          \includegraphics[width=1\linewidth]{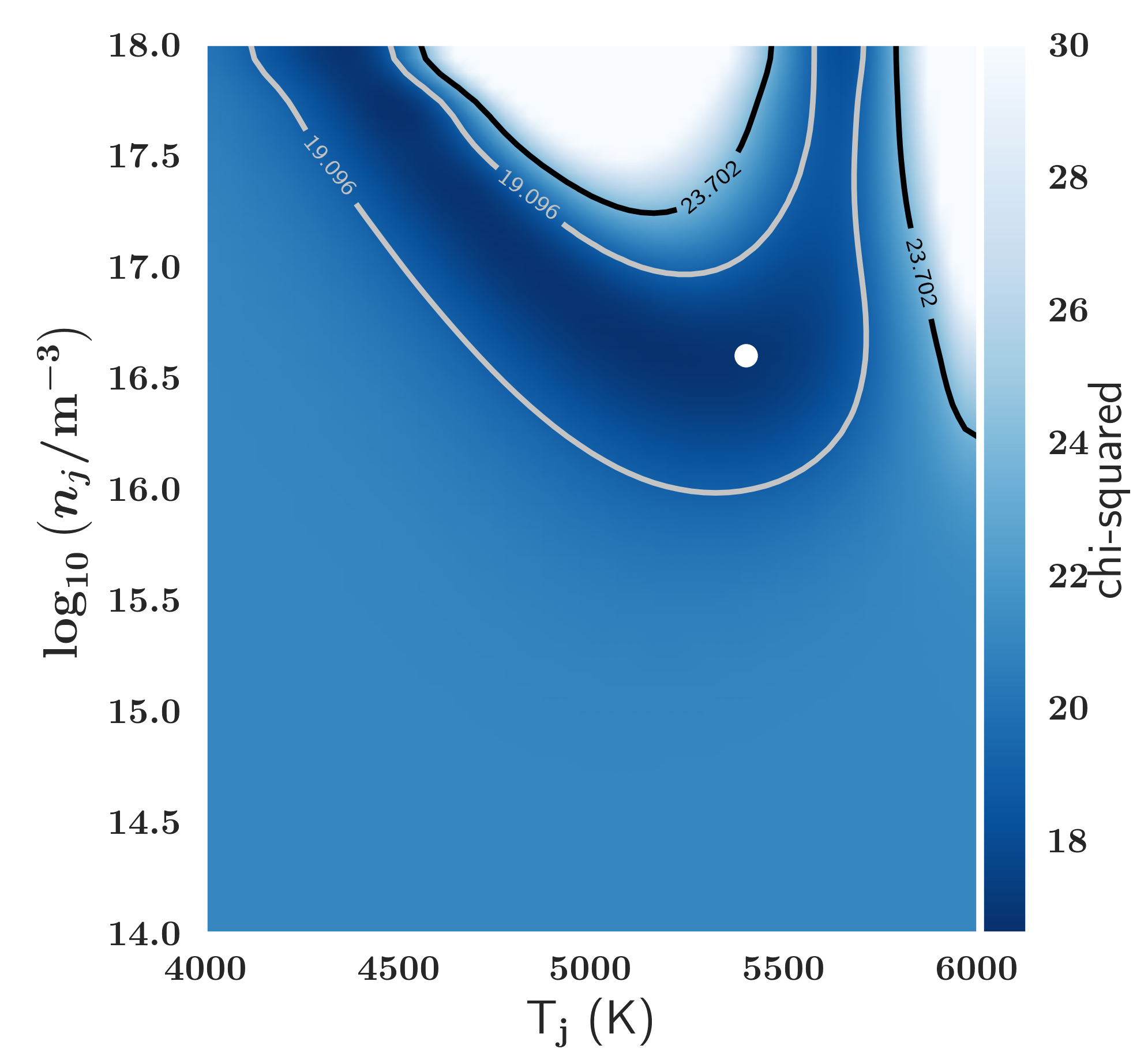}
        \caption{HD 46703}\label{fig:chisq_hd}
    \end{subfigure}%
    \begin{subfigure}[b]{.4\linewidth}
        \centering\large 
          \includegraphics[width=1\linewidth]{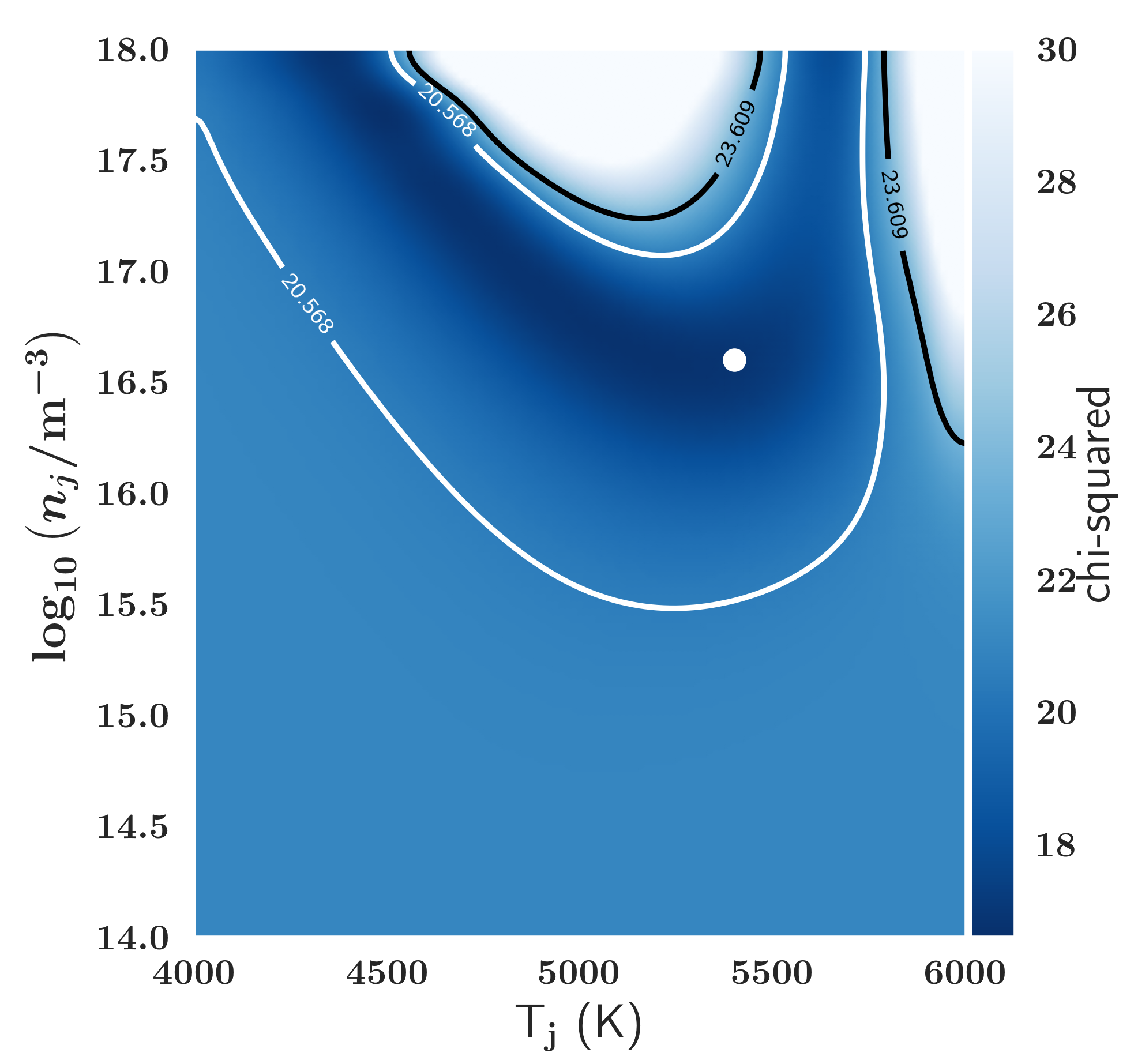}
        \caption{HP Lyr}\label{fig:chisq_hplyr}
    \end{subfigure}%
    
    \begin{subfigure}[b]{.4\linewidth}
        \centering\large 
          \includegraphics[width=1\linewidth]{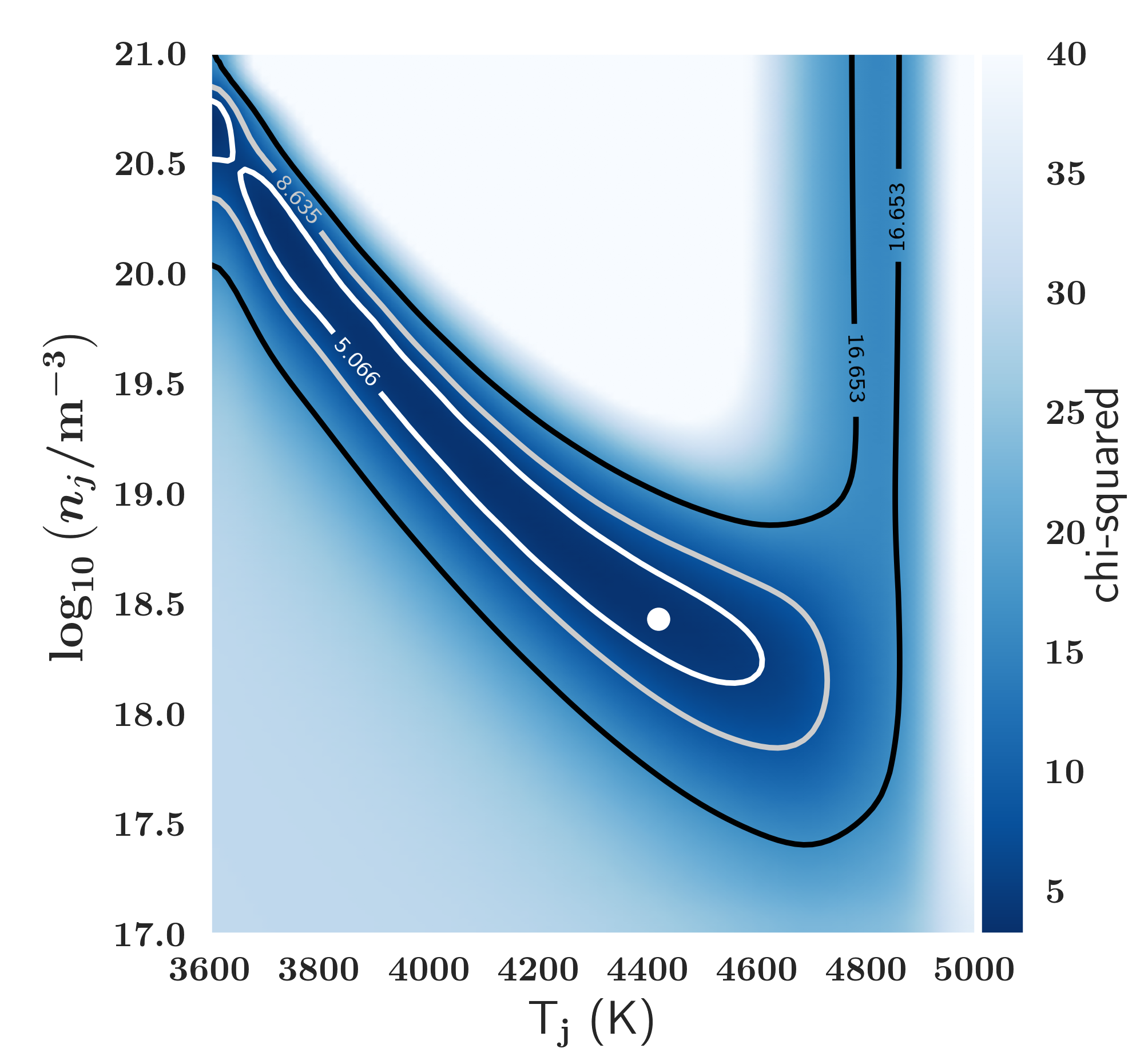}
        \caption{TW Cam}\label{fig:chisq_twcam}
    \end{subfigure}
    \caption{Two-dimensional reduced chi-squared distribution for the grid of jet densities $n_j$ and temperatures T$_j$ for each object in the sample. The white dot indicates the location of the best-fitting model. The contours represent the $1\sigma$, $2\sigma$, and $3\sigma$ intervals.} \label{fig:chisq}
\end{figure*}

\begin{figure*}
  \centering
    \begin{subfigure}[b]{.8\linewidth}
        \centering\large 
          \includegraphics[width=1\linewidth]{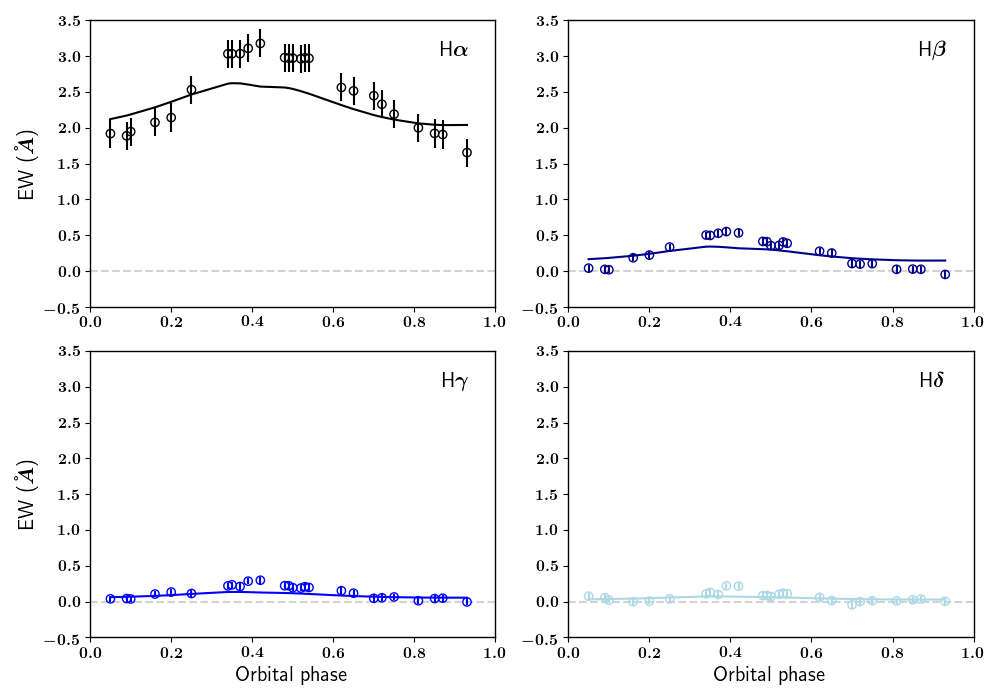}
        \caption{89\,Her}\label{fig:ewfit_89}
    \end{subfigure}%
    
    \begin{subfigure}[b]{.8\linewidth}
        \centering\large 
          \includegraphics[width=1\linewidth]{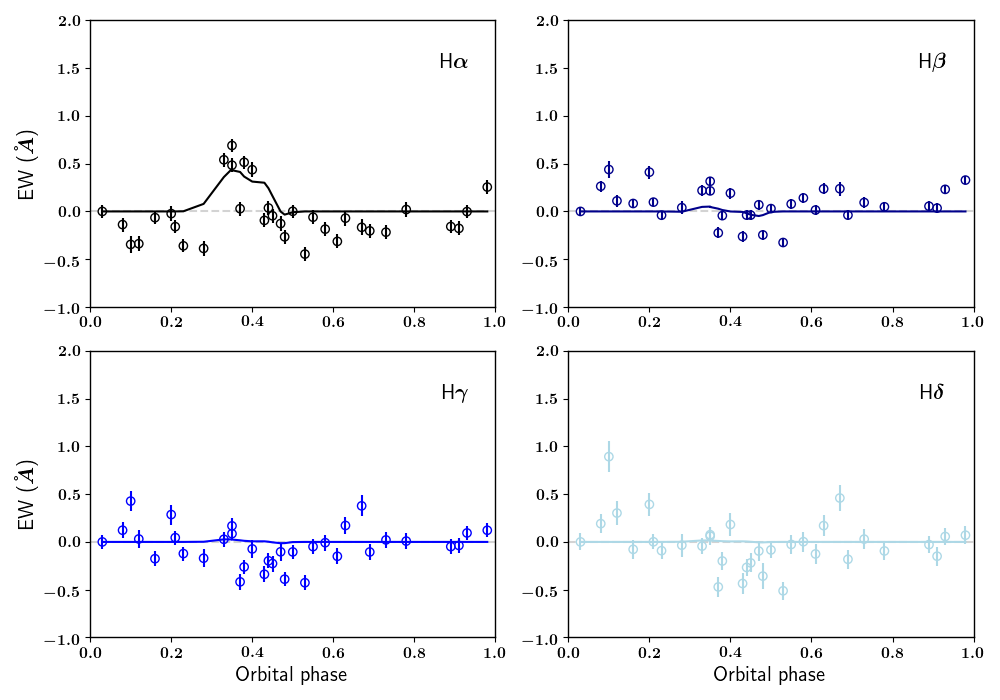}
        \caption{HD 46703}\label{fig:ewfit_hd}
    \end{subfigure}%
    \caption{EW of the absorption by the jet for 89\,Her and HD\,46703. The panels show the EW in \halpha, \hbeta, \hgamma, and \hdelta. The circles and full lines are the measured EWs of the jet absorption feature for the observations and the best-fitting model, respectively. }  \label{fig:ewfit}
\end{figure*}

\begin{figure*}
  \centering    
    
    \begin{subfigure}[b]{.8\linewidth}
        \centering\large 
          \includegraphics[width=1\linewidth]{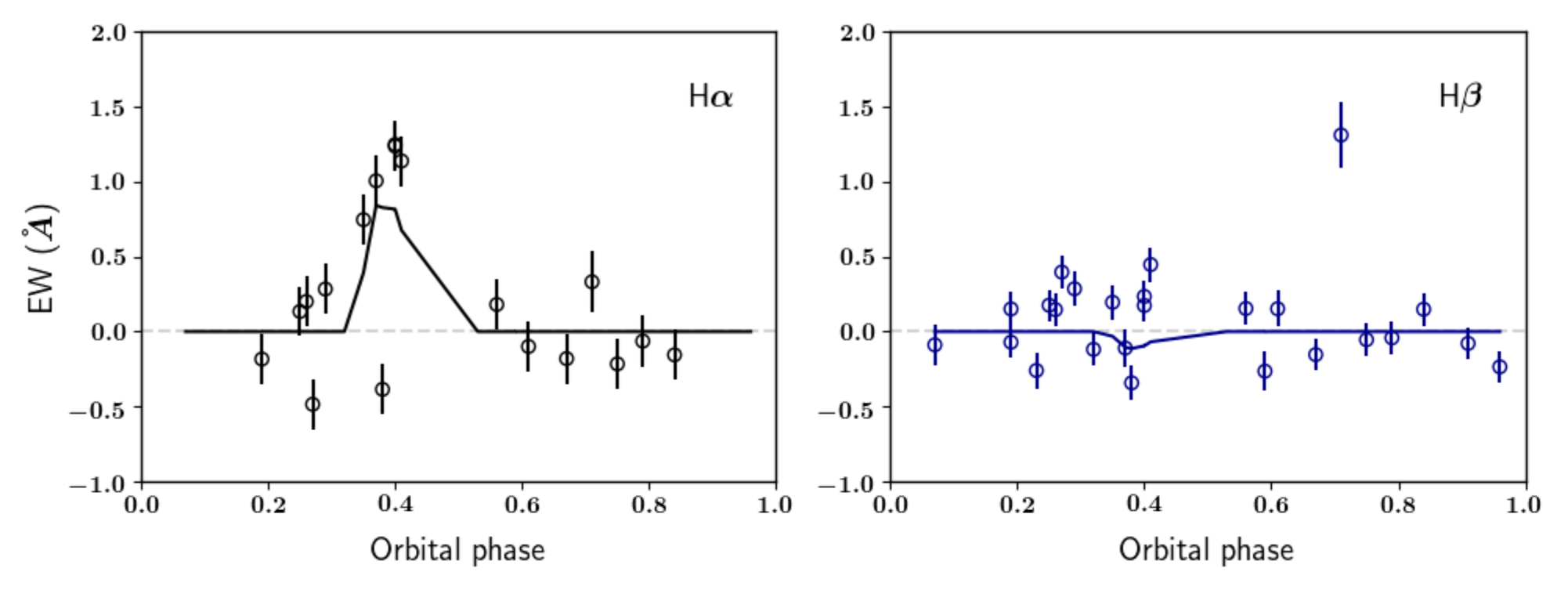}
        \caption{EP Lyr}\label{fig:ewfit_eplyr}
    \end{subfigure}%
    
    \begin{subfigure}[b]{.8\linewidth}
        \centering\large 
          \includegraphics[width=1\linewidth]{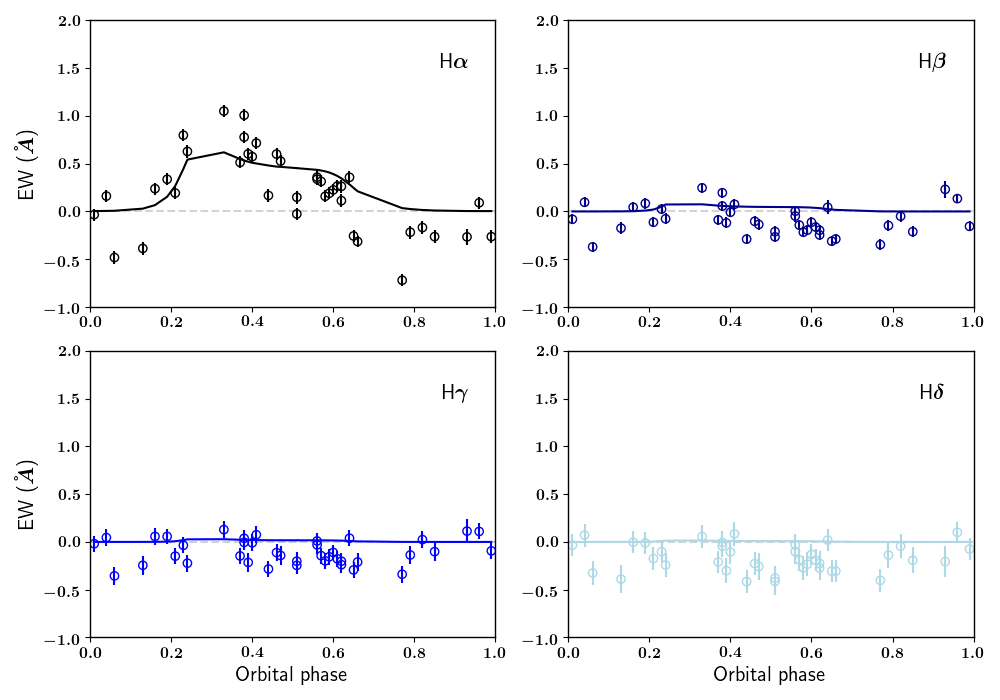}
        \caption{HP Lyr}\label{fig:ewfit_hplyr}
    \end{subfigure}%
    
    \begin{subfigure}[b]{.8\linewidth}
        \centering\large 
          \includegraphics[width=1\linewidth]{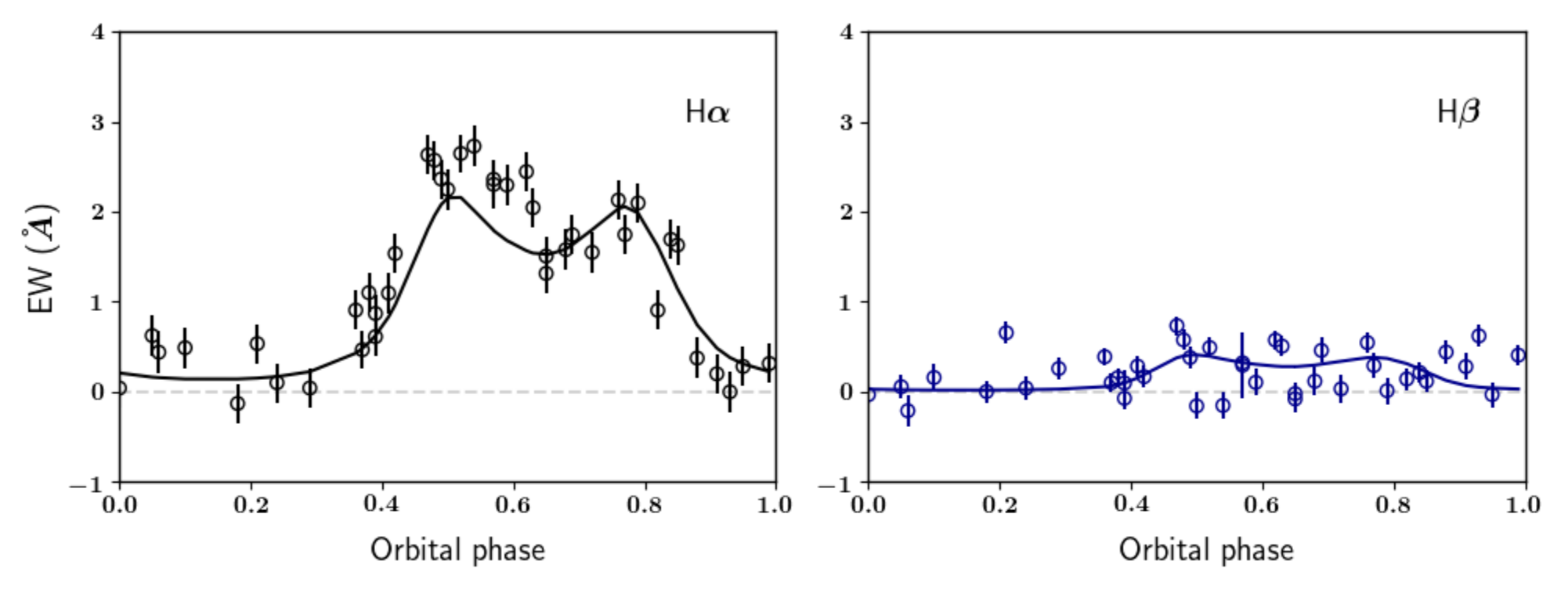}
        \caption{TW Cam}\label{fig:ewfit_twcam}
    \end{subfigure}
    \caption{Similar as figure~\ref{fig:ewfit}, but for EP\,Lr, HP\,Lyr, and TW\,Cam. For EP\,Lyr and TW\,Cam, we excluded the \hgamma\, and \hdelta\, line since the uncertainties in these two lines are too high. } \label{fig:ewfit2}
\end{figure*}

\bsp	
\label{lastpage}
\end{document}